\documentclass[11pt,twoside,english]{article}

\usepackage{setspace}

\usepackage{geometry}
\usepackage{indentfirst}

\usepackage{graphicx}

\graphicspath{{../figures/}}

\usepackage{amssymb}
\usepackage{amsbsy}
\usepackage{amsmath}

\usepackage{lscape}
\usepackage{natbib}

\usepackage{fancyheadings}

\title{Bayesian Learning and Predictability in\\ 
a Stochastic Nonlinear Dynamical  Model}
\author{John Parslow$^a$, Noel Cressie$^b$, Edward P. Campbell$^c$
\footnote{Corresponding author. E-mail address: eddy.campbell@csiro.au},\\ 
Emlyn Jones$^a$, Lawrence Murray$^c$\\
~~~\\
$^a$\textit{CSIRO Computational and Simulation Science- Marine and}\\
 \textit{Atmospheric Research, GPO Box 1538, Hobart, TAS 7001, Australia}\\
$^b$\textit{University of Wollongong and Department of Statistics, The Ohio State University,}\\ 
\textit{1958  Neil Avenue, 404 Cockins Hall, Columbus, OH  43210-1247, USA}\\
$^c$\textit{CSIRO Computational and Simulation Science- Mathematics,}\\
\textit{Informatics and Statistics, Private Bag 5, WA 6913, Australia}
}
\date{}

\begin{document}

\pagestyle{fancyplain}
\rhead[]{In press at \emph{Ecological Applications}}
\lhead[In press at \emph{Ecological Applications}]{}

\maketitle

\begin{abstract}
Bayesian inference methods are applied within a Bayesian hierarchical
modelling framework to the problems of joint state and parameter
estimation, and of state forecasting. We explore and demonstrate the
ideas in the context of a simple nonlinear marine
biogeochemical model. A novel approach is proposed to the formulation
of the stochastic process model, in which ecophysiological properties
of plankton communities are represented by autoregressive stochastic
processes. This approach captures the effects of changes in plankton
communities over time, and it allows the incorporation of literature
metadata on individual species into prior distributions for process
model parameters. The approach is applied to a case study at Ocean
Station Papa, using Particle Markov chain Monte Carlo computational techniques. The
results suggest that, by drawing on objective prior information, it is
possible to extract useful information about model state and a subset
of parameters, and even to make useful long-term forecasts, based on
sparse and noisy observations.
\end{abstract}


\setlength{\parindent}{0.5cm}

\section{Introduction}\label{sec:intro}

The last century has seen major advances in the ecological and earth
sciences, both in the development of theoretical understanding,
encapsulated in mechanistic process models, and in the development of
sophisticated statistical theories and models for the interpretation
and analysis of observations. However, as \cite{Berliner2003} has
pointed out, until recently the development of process models and the
statistical analysis of observations have occurred in parallel and
somewhat at arms length. Over the last two decades, there has been
increasing effort devoted to the integration of observations and
process models, so that model--data comparison and data assimilation
are now key research topics.

There are a number of drivers for this increased emphasis on the
integration of models and observations. The scientific community
increasingly insists on the use of more objective and quantitative
measures  or metrics to evaluate model predictions against
observations \citep[e.g.][]{Allen2007}. But ecological and earth system
models are increasingly used for practical purposes, from short-term
environmental forecasting to local issues of pollution, conservation
and renewable resources, to global issues of climate change. Users of
model outputs would like more accurate predictions and increasingly
demand formal assessments of the uncertainty in model predictions, to
inform decision-making and risk-management.

Techniques for the integration of models and observations are intended
to quantify model performance and allow intercomparison of
alternative  models, to improve performance or skill in model
predictions, and to provide error estimates or confidence/credible intervals
around  those predictions. Errors enter into an integrated model--data
system from at least three sources. First, there are errors in the process
of making observations, which typically provide a distorted and/or
fragmented glimpse of the underlying reality. One consequence is that
we do not know the exact state of the system when we initialise
dynamic models. Second, process models make simplifying assumptions and
approximations, so that model simulations cannot be expected to
reproduce reality exactly. Many ecological and earth system models
are dynamic models, predicting the evolution of system trajectories
over time, and model errors are typically stochastic, leading to
divergence of simulated trajectories over time. Finally, process
models typically incorporate a number of parameters, assumed constant
over time, whose values are uncertain.

The term ``data assimilation'' has been used broadly to describe
model--data integration \citep[e.g.][]{Gregg2008, LuoY2011}. In
practice, approaches and applications have tended to fall into one of
two categories. In the first, attention has focused on the estimation
of uncertain parameters in deterministic process models
\citep[e.g.][]{Matear1995}. Parameters are often estimated by
minimizing some kind of cost function based on model--data mismatches,
typically a sum-of-squared errors. In some cases, the
cost function is constructed and interpreted as a negative
log-likelihood based on a formal error model but, in other cases, the
cost function is ad hoc. The second class of applications typically
involves short-term environmental forecasting or hindcasting, where
errors are believed to be dominated by uncertainty about the true
value of the system state. Sequential data assimilation techniques
are used to update estimates of the state based on current or recent
observations. In these approaches, there tends to be a strong emphasis
on building realistic observation models, while the stochastic model
error is often modelled as simple additive white noise and adjusted
to  achieve convergence of the assimilation procedure. Very
sophisticated data assimilation schemes are now widely adopted and
routinely used in weather and ocean forecasting.

The last decade especially has seen increasing advocacy of Bayesian
approaches to data assimilation \citep[e.g.][]{Link2001, Berliner2003,
Calder2003, Cressie2009, Zobitz2011}. Bayesian methods  typically
yield posterior distributions for the inferred state and parameters,
most often summarised using large samples from  these
distributions. These can be particularly useful in
applied  contexts, where users may be interested in the probability
distribution  of performance measures derived from model
predictions. A key attraction of the Bayesian approach is
its  ability to formally incorporate prior information about models
and parameters.  Given that the rationale for using mechanistic,
process-based  models is that they build on prior scientific knowledge
about  the structure and function of system components, it makes
sense to  use methods that allow this knowledge to be formally
represented  in model--data comparisons. It is of course possible to
use the  Bayesian formalism, while discounting or ignoring prior
information,  through uninformative priors or empirical Bayes
methods. In these  cases, Bayesian methods can generally be shown to
be  equivalent to classical methods  \citep[e.g.][]{VerHoef1996a,
Cressie2009}.

Within the broader Bayesian tradition, Bayesian Hierarchical Modelling
(BHM) offers a particularly attractive framework for the integration
of mechanistic process models and observations. BHM provides a
consistent, formal probabilistic framework combining error or
uncertainty in model parameters, model state, model processes and
observations \citep{Wikle2003, Berliner2003, Cressie2009}. This
framework encourages the modeller to think carefully and
systematically about the approximations and assumptions involved in
process model formulation, about the observation process and the
relationship between model state variables and observations, and about
the relationship between model parameters and independent prior
knowledge. One can think of BHM not just as an integration of models
and data, but as a deep integration of mechanistic and statistical
modelling; \cite{Berliner2003} describes this as
``physical-statistical'' modelling.

The last decade has seen a rapid growth of Bayesian applications in
ecology  and the earth sciences, ranging from population dynamics and
dispersal \citep[e.g.][]{Link2001, Calder2003, Wikle2003,
  Clark2004a, Clark2006, Barber2007, Hooten2007a}  to plant ecology and terrestrial
surface  fluxes \citep[e.g.][]{Ogle2004, Baker2006, Sacks2006,
Xu2006, Zobitz2007, Zobitz2008} to ocean circulation  and climate 
\citep[e.g.][]{Berliner2000, Berliner2003}. Encouragingly,
Bayesian approaches  are now widely and successfully used for stock
assessment and fisheries management \citep{Maunder2004}. 

In this paper, we focus on the application of Bayesian methods,
specifically BHM,
to aquatic  biogeochemical (BGC)/ecological models. Model--data integration
in this  field has paralleled the broader trajectory outlined
above. Earlier studies focused on the problem of parameter estimation
in deterministic models \citep{Matear1995}. Over the last
decade, and  following developments in data assimilation into physical
ocean  circulation models, there has been considerable progress in
implementing  sequential data assimilation techniques for state
estimation in  3-D biogeochemical models \citep{Gregg2008}. Examples
of Bayesian approaches in this area fall into two
streams.  The first uses a Bayesian approach to obtain
posteriors for parameters and state estimation in (effectively)
deterministic eutrophication models \citep{Arhonditsis2008,
 Arhonditsis2007, Zhang2009}. The second, in contrast, uses sequential Bayesian assimilation to obtain
posteriors  for current and forecast state in stochastic models in
which  the underlying parameters are assumed constant and known 
\citep{Dowd2003, Dowd2006, Dowd2007}. More
recently, \cite{Dowd2011} has extended this work to obtain joint
posteriors for the state and a subset of parameters. These examples all
embed the ecological dynamics physically within a 0-D box model
setting, but \cite{Mattern2010} extend this to a 1-D setting.

The study presented here aims to build on previous work by using the
BHM probabilistic framework to underpin enhancements in several areas:

\begin{enumerate}

  \item The process models used here include stochastic errors in a way
  that  accounts for key simplifying approximations made in replacing
  communities  of species by a single biomass variable. These
  approximations  are widely used in ecological and biogeochemical
  models,  and the approach seems likely to find broader application.

  \item Our approach also allows prior distributions for model
    parameters  to be more directly and objectively related to prior
    information  obtained from field and laboratory studies, and
    from in literature meta--data. This prior information makes a
    valuable  contribution to state estimation and forecasting in the
    application  considered here, where observations are severely
    limited. 

  \item The process model has been modified to include a diagnostic
    variable, Chlorophyll a, to support a simpler and more rigorous
    observation model.

  \item Bayesian inference in nonlinear problems is generally
    analytically  intractable, and computationally intensive
    simulation-based  methods, such as Markov chain Monte Carlo, are
    used  to obtain large random samples from the posterior. Our
    study  exploits new methods for Bayesian inference
    \citep{Andrieu2010} to derive a joint posterior for parameters
    and state in nonlinear dynamical models. This allows us to
    simultaneously address problems of parameter estimation, state
    estimation, short-term forecasting and long-term projections in a
    unified probabilistic framework.

\end{enumerate}

The remainder of this paper is organised as follows. In Section \ref{sec:method}, we
provide  a brief introduction to BHM and its application to dynamical
state-space models. Section \ref{sec:BHMapp} presents  a reformulation of a
conventional  deterministic model as a stochastic process model within
the  BHM framework. Uncertainty in the parameters is captured through
a collection of time-varying stochastic processes. In Section \ref{sec:learning}, we
provide a case study of this generic  model  applied to a
time  series of observations at Ocean Station Papa.  Bayesian inference
procedures   are used to extract  information in the form of
posteriors for state and parameters  from a set of observations that are sparse
and  patchy in time, and include only a subset of state
variables. Twin  experiments are used to test the performance and
consistency  of the inference procedures, and to draw some preliminary
conclusions about the effect of observation intensity on
posteriors. Section \ref{sec:DandC}  discusses the results obtained in
the  context of the enhancements listed above, and we make some
observations  about the strengths and weaknesses of this approach for
marine  biogeochemical modelling, and ecological modelling more
broadly. This is followed by mathematical, statistical and computing appendices.

\section{General Methodology}\label{sec:method}

\subsection{Bayesian Hierarchical Models (BHMs)}\label{sec:BHM}

The physical-statistical models described by \citet{Berliner2003}, formulated as  BHMs, are  models that explicitly represent three sources of uncertainty:

\begin{enumerate} 
	\item Data model: Expresses uncertainty arising from observations  subject to measurement error and bias.
	\item Process model: Expresses uncertainty arising from
          scientific (here, biophysical) processes that are not
          completely understood or they are approximated.
	\item Parameter (prior) model: Expresses uncertainty arising from parameters not known exactly. 
\end{enumerate}

BHMs are probabilistic models, constructed from conditional
probability distributions. The data are treated as conditional on the
process and some parameters, and the process is treated as conditional
on other parameters.  Hence, the three components, data, processes and
parameters can be thought of as hierarchical levels in a chain of
conditional dependence, which we now formalise.

Let the data (observations), process(es) and  parameters be
represented by the vectors ${\bf{Y}}$, ${\bf{W}}$   and
$\boldsymbol{\theta}$, respectively.  In some models, the process has a
continuous index in time or space; for the purpose of computations it
is enough to consider $\mathbf{W}$ as a high-dimensional vector. The joint
uncertainty is denoted $[{\bf{Y}}, {\bf{W}}, \boldsymbol{\theta} ]$,
where the notation $[\bf{A}]$  represents ``the probability distribution of \textbf{A}.'' It makes sense to  partition the parameters into biophysical parameters  and so-called statistical parameters arising from the observation process.  Therefore, we write $\boldsymbol{\theta}  =\{\boldsymbol{\theta}_{\bf{Y}},\boldsymbol{\theta}_{\bf{W}}\}$.

Applying the rules of conditional probability, we can factorise the joint probability distribution as:
\begin{equation}
	[{\bf{Y}},{\bf{W}},\boldsymbol{\theta} ] = [{\bf{Y}} \vert {\bf{W}},\boldsymbol{\theta}_{\bf{Y}}, \boldsymbol{\theta}_{\bf{W}}][{\bf{W}}, \boldsymbol{\theta}_{\bf{Y}},\boldsymbol{\theta}_{\bf{W}}],
\label{eq:YWtheta1}
\end{equation}
\noindent where $[\bf{A}|\bf{B}]$ denotes ``the conditional probability of \textbf{A} given \textbf{B}''. Repeating this for the second component of (\ref{eq:YWtheta1}), we find:
\begin{equation}
	[{\bf{Y}},{\bf{W}},\boldsymbol{\theta} ] = [{\bf{Y}}\vert {\bf{W}},\boldsymbol{\theta}_{\bf{Y}}, \boldsymbol{\theta}_{\bf{W}}][{\bf{W}}\vert \boldsymbol{\theta}_{\bf{Y}},\boldsymbol{\theta}_{\bf{W}}] [\boldsymbol{\theta}_{\bf{Y}},\boldsymbol{\theta}_{\bf{W}}]\,.
\label{eq:YWtheta2}
\end{equation}	

\noindent The components of (\ref{eq:YWtheta2})   may be simplified a
little by noting that the biophysical  parameters,
$\boldsymbol{\theta}_{\bf{W}}$, are not needed  in the data model when
we also condition on the process;  similarly, the statistical
parameters,  $\boldsymbol{\theta}_{\bf{Y}}$, are not needed in the second component when we also condition on the  biophysical parameters.  Hence, we obtain:
\begin{equation}
	[{\bf{Y}},{\bf{W}},\boldsymbol{\theta} ] = 
[{\bf{Y}}\vert {\bf{W}},\boldsymbol{\theta}_{\bf{Y}}]
[{\bf{W}}\vert\boldsymbol{\theta}_{\bf{W}}]
[\boldsymbol{\theta}_{\bf{Y}},\boldsymbol{\theta}_{\bf{W}}]\,.
\label{eq:YWtheta3}
\end{equation}

We see that the three probability distributions on  the right-hand
side correspond to the BHM hierarchy of  sources of uncertainty
identified above, representing  a data model, a (stochastic)
biophysical process  model, and a parameter model, respectively.  The parameter model is often referred to as the  \textit{prior distribution}.

Of key interest is how one can make inferences about the  unobserved
process state ${\bf{W}}$ and the parameters  $\boldsymbol{\theta}$, given the observations  ${\bf{Y}}$ on the biogeochemical process. Appealing to Bayes'  Theorem (e.g., Cox and Hinkley, 1986, pp.\ 365-367), we may  write:
\begin{equation} 
	[{\bf{W}},\boldsymbol{\theta}\vert {\bf{Y}}] 
\propto [{\bf{Y}}\vert {\bf{W}},\boldsymbol{\theta}_{\bf{Y}} ]
[{\bf{W}}\vert\boldsymbol{\theta}_{\bf{W}} ][\boldsymbol{\theta}_{\bf{Y}}, \boldsymbol{\theta}_{\bf{W}}]\,,
	\label{eq:PthetaD} 
\end{equation} 

\noindent where the constant of proportionality is a function  of ${\bf{Y}}$ only and guarantees that the right-hand side of  (\ref{eq:PthetaD}) is a proper joint probability  distribution. This so-called \textit{posterior distribution}  is proportional to the product of the three levels of the  BHM (data model, process model, parameter model) that we  have developed above.  We return later in this section to  the issue of making inferences based on (\ref{eq:PthetaD}).

The use of the three levels of conditional probability models via  Bayes' Theorem to learn from data is precisely the BHM framework we alluded to at the beginning  of this section.  Examples of its use have been growing in  the last decade.  It was introduced in a climate-modelling  and climate-prediction context by \citet{Berliner2000},  in an introductory geophysical context by  \citet{Berliner2003}, and in an ecological context by  \citet{Wikle2003}; see also the review by \citet{Cressie2009}.

\subsection{A State Space Representation}\label{sec:SSR}
We are interested here in the application of BHM to dynamical systems,
in which the state evolves as a function of time (discrete or
continuous), and the data are collected by sampling (potentially
irregularly and  coarsely) in time, whilst the process evolves at a
relatively fine time step.  We write the time-evolving process
${\bf{W}}$ as $({\bf{W}}_0, {\bf{W}}_1,\dots ,{\bf{W}}_T)$ with
corresponding observations $({\bf{Y}}_1,\dots ,{\bf{Y}}_T)$ taken after the
initial value of the process $\bf{W}_0$.  We use subscript $t$  to
index time, such that ${\bf{W}}_t$ is coincident with  ${\bf{Y}}_t$,
for $t=1, \dots, T$. A graphical depiction of the dependencies is
shown in Figure~\ref{fig:statespace} below.

\begin{figure}[htp]
\centering
\includegraphics[scale=1.4]{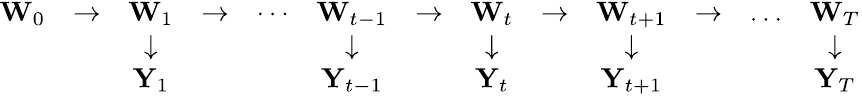}
\caption{A graphical representation for the process ${\bf{W}}$ and observations ${\bf{Y}}$}
\label{fig:statespace}
\end{figure}

\noindent We remark that, in practice, observations will be missing at some times, which the BHM framework can readily handle.

We henceforth assume that the forward evolution of the process
${\mathbf{W}}$ depends only on the current state; that is, ${\bf{W}}$ is a
Markov-process model described by  $[{\bf{W}}_t\vert
\mathbf{W}_{t-1},\boldsymbol{\theta}_{\bf{W}}]$, for $t=1,\cdots,T$.
This form of conditional independence implies that ${[\bf{W} \vert
\boldsymbol{\theta}_{\bf{W}}]} =  \prod_{t=1}^{T}[{\bf{W}}_t \vert
{\bf{W}}_{t-1}, \boldsymbol{\theta}_{\bf{W}}]$.  Further, observations
at time $t$ are assumed to be independent of observations at other
times, conditional on the state ${\bf{W}}_t$. Thus, the data model has
the form, ${[\bf{Y} \vert \bf{W}, \boldsymbol{\theta}_{\bf{Y}}]} =
\prod_{t=1}^{T}[{\bf{Y}}_t\vert
{\bf{W}}_{t},\boldsymbol{\theta}_{\bf{Y}}]$.

\subsection{Statistical Inference}\label{sec:inference}
The focus of our statistical inference is the calculation of  the
posterior distribution described by Equation  (\ref{eq:PthetaD}),
which is rarely amenable to analytic solutions. As a result, 
modern Bayesian inference has harnessed efficient
algorithms deployed on contemporary computing architectures to
simulate samples from the posterior distribution. Statistics
calculated for these samples, such as means and quantiles, can be
shown to converge to the appropriate quantities for the posterior
distribution \citep{Tierney1994}.

Suppose for instance that we are interested in estimating some
function $g(\cdot)$ of the state and parameters. We obtain a simulated
sample $\{({\bf{W}}^{(\ell )},\boldsymbol{\theta}^{(\ell )})\colon\ell
=1,\dots ,L\}$  from the posterior distribution
$[{\bf{W}},\boldsymbol{\theta}\vert{\bf{Y}}]$,  and we use the
transformed sample $ \{g({\bf{W}}^{(\ell )},\boldsymbol{\theta}^{(\ell
  )})\colon\ell=1,\dots ,L\} $ to calculate summary statistics. For example, we can
estimate the mean as: 

 \[ \widehat{E}(g({\bf{W}},\boldsymbol{\theta})\vert{\bf{Y}}) \equiv (1/L)\sum^L_{\ell =1} g({\bf{W}}^{(\ell )},\boldsymbol{\theta}^{(\ell )}), \]

\noindent so sampling from the posterior distribution over
states and parameters is key to the success of Bayesian hierarchical
modelling in this context. The computational approach adopted must also be able to
cope with the nonlinear behavior of the process model, noting that the
state transition density function is not available in closed form.

Particle Markov chain Monte Carlo (PMCMC) was developed for exactly this
situation, and so we have applied it in our case study. In particular, we use
the particle marginal Metropolis-Hastings (PMMH) sampler~\citep{Andrieu2010},
which we have previously applied successfully to a simple Lotka-Volterra type
model~\citep{Jones2010}.  Details of PMMH are given in Appendix \ref{app:PMMH}

\section{Reformulating a Marine BGC Model as a BHM}\label{sec:BHMapp}

A general description of the BHM framework and its use for scientific inference was given in Section~\ref{sec:method}.  We now show how these ideas can be applied in a marine BGC setting.

\subsection{The process model}\label{sec:procmodel}

Recall from Section~\ref{sec:method} that the  biogeochemical process model is at the second level of the BHM hierarchy.   We present the model first in terms of a deterministic model,  and then we derive a stochastic version of it.

\subsubsection{A deterministic biogeochemical process model}\label{sec:BGC}

One of the advantages of the BHM framework is that it  allows us to build on existing  scientific understanding, typically incorporated in  deterministic process models.  We can draw here on a  long and rich history of (deterministic) marine  BGC models that describe  the cycling of nutrients (e.g. nitrogen) and/or carbon through  living and nonliving organic and inorganic compartments,  in simplified marine ecosystems.  Open-ocean models  typically deal only with pelagic planktonic systems,  while coastal models may deal with coupled pelagic-benthic  systems.  In this article, we deal with the simpler case of  pelagic models.

In the general case, the state variables in marine BGC models are expressed as component  concentrations (mass per unit volume) as functions  of space \textbf{x} and time $t$. These components  are subject to physical transport (advection and mixing),  as well as local biological and chemical reactions. If  \textbf{c}(\textbf{x},$t$) is a vector of state variables,  we can write the general reaction-transport equation as: 

\begin{equation} 
	\frac{\partial{\bf{c}}}{\partial{t}}={\bf R}({\bf c},{\bf x},t) + {\bf T}({\bf c},{\bf x},t)\,,
\label{eq:partialc}
\end{equation}

\noindent where \textbf{R} represents local biological and  chemical
reactions, and \textbf{T} is a transport  operator; see Appendix \ref{app:sourcesink} 
for the specific form of (\ref{eq:partialc}) used in the case study in
section \ref{sec:learning}.  In this paper, we consider the highly
simplified physical setting of a mixed-layer one-box model, and for
the moment we ignore the transport operator and  focus on the
local reactions \textbf{R}. This setting allows us to formulate a BHM
most clearly. However, we do include a simple transport term to
account for vertical mixing in the case
study and this is presented in Appendix \ref{app:sourcesink}.

Pelagic planktonic ecosystems are complex systems  that involve many species of phytoplankton and zooplankton,  multiple (potentially limiting) nutrients, and dissolved  and particulate organic matter pools comprised of  complex mixtures. All models of these systems require  simplifying approximations, and the level of  detail varies across models and depends  on the purpose of the model.  Model detail and  complexity have tended to increase over the last decade,  as scientific understanding and computational  power have increased.  However, this in turn has led to concern about  the identifiability of complex models with many uncertain  parameters \citep{Hood2006}.

We have chosen a relatively simple, classic NPZD model  formulation, which represents the cycling of a limiting  nutrient (nitrogen) through four compartments: dissolved inorganic nitrogen or DIN ($N$), phytoplankton nitrogen ($P$),  zooplankton nitrogen ($Z$), and detrital nitrogen ($D$).  We can write the equations for the local rate of change  of the state variables as:

\begin{equation}
	\frac{dP}{dt}=g\cdot P-gr\cdot Z, \label{eq:dPdt}
\end{equation}

\begin{equation}
	\frac{dZ}{dt}=E_Z\cdot gr\cdot Z - m\cdot Z , \label{eq:dZdt}
\end{equation}

\begin{equation}
	\frac{dD}{dt}=(1 - E_Z)\cdot f_D\cdot gr\cdot Z + m\cdot Z - r\cdot D, \label{eq:dDdt}
\end{equation}

\begin{equation}
	\frac{dN}{dt}=-g\cdot P + (1 - E_Z)\cdot (1- f_D)\cdot gr\cdot Z + r\cdot D .\label{eq:dNdt}
\end{equation}

\noindent Notice that
$\frac{dP}{dt}+\frac{dZ}{dt}+\frac{dD}{dt}+\frac{dN}{dt}=0$ which is a
consequence of ``mass balance'' in the currency of nitrogen. In
(\ref{eq:dPdt}) - (\ref{eq:dNdt}), $g$ is the phytoplankton specific
growth rate  (per day, or d$^{-1}$), $gr$ is the zooplankton specific
grazing  rate (mg $P$ grazed per mg $Z$ d$^{-1}$), $m$ is the
zooplankton specific mortality rate (d$^{-1}$),  and $r$  is the
specific breakdown rate of detritus (d$^{-1}$).  A fraction $E_Z$ of
zooplankton ingestion is converted  to zooplankton growth and, of the
remainder, a fraction  $f_D$ is allocated to detritus, with the rest
released  as dissolved inorganic nitrogen, $N$.  The fractions, $E_Z$
and $f_D$, are treated as constant, independent of ingestion
rates. This is a common simplifying assumption in biogeochemical
models \citep[e.g.,][]{Wild-Allen2010}.  

The process rates $g$, $gr$, $m$, and $r$ are all  functions of state
variables and/or exogenous forcing variables, and hence they are
functions of time. As we shall see below, a multiplicative temperature
correction $Tc$ is applied to all rate processes; to define $Tc$, we
use a so-called ``$Q_{10}$ formulation'' for dependence on temperature $T$:

\begin{equation}
	Tc = Q_{10}^{(T - T_{ref})/10}\,, \label{eq:Tc}
\end{equation}

\noindent Notice that $T$ depends on time and, hence, so does $Tc$, where $T_{ref}$ is a reference temperature, and  $Q_{10}$ is a prescribed parameter.

We use a flexible formulation for the dependence of zooplankton's grazing rate on phytoplankton concentration  (zooplankton functional response): 

\begin{equation}
	gr = \frac{Tc\cdot I_Z\cdot A^\upsilon}{(1 + A^\upsilon )}\,, \label{eq:gr}
\end{equation}

\noindent where $\upsilon$ is a given power; the relative availability of phytoplankton $A$ is

\begin{equation}
	A = \frac{Cl_Z\cdot P}{I_Z}\,, \label{eq:A}
\end{equation}

\noindent where $A$ depends on time because $P$ does. In (\ref{eq:A}), $I_Z$ is the maximum zooplankton ingestion  rate (mg $P$ per mg $Z$  d$^{-1}$); and $Cl_Z$ is the  maximum clearance rate (volume in m$^3$ swept  clear per mg $Z$ d$^{-1}$). Both are constant in the deterministic formulation.  This is a  standard rectangular hyperbola or Type-2  functional response \citep{Holling1965} when $\upsilon = 1$, and a Type-3 sigmoid  functional response when $\upsilon >1$.

We follow \citet{STEELE1976} and \citet{STEELE1992} in adopting a time-dependent quadratic formulation for zooplankton mortality:

\begin{equation}
	m = Tc\cdot m_Q\cdot Z\,, \label{eq:m} 
\end{equation}

\noindent where the constant quadratic mortality rate $m_Q$ has units of (mg $Z$ m$^{-3}$)$^{-1}$ d$^{-1}$. The detrital remineralization rate is assumed to depend only on temperature (which is time dependent):

\begin{equation}
	r = Tc\cdot r_D\,, \label{eq:r}
\end{equation}

\noindent where the constant parameter $r_D$ prescribes the rate  at the reference temperature and has units of d$^{-1}$.

Finally, the phytoplankton specific growth rate $g$  depends on
temperature $T$, available light  or irradiance $E$ (see Appendix \ref{app:lightmodel}
)  and dissolved inorganic nitrogen $N$.  The submodel given below for $g$ is somewhat more elaborate than the submodels  used for the other rate processes.  We shall see that it predicts changes in  phytoplankton composition (nitrogen:carbon ratio and  chlorophyll-a:carbon ratio) as well as the phytoplankton specific  growth rate, as phytoplankton adapt to changes in  available light and nutrients.

In the BHM framework, we are encouraged to pay careful  attention to the relationship between process model  variables and what we can observe.  For example, the process  model predicts phytoplankton biomass $P$ in the currency of mg $N$ m$^{-3}$, but we typically measure  phytoplankton as a pigment (mg $Chla$ m$^{-3}$). The submodel given in the following paragraphs allows us to relate these chlorophyll observations  ($Chla$) more rigorously to the state variable $P$. Our formulation  represents a variant on models proposed by \citet{Geider1998}, and details of our derivation are given in Appendix \ref{app:Pg_derivation}.

The phytoplankton specific growth rate $g$ is expressed in terms of $g^{max}$ (in units of d$^{-1}$), a constant maximum specific growth rate at the reference temperature, $T_{ref}$, a light-limitation term $h_E$, and a  nutrient-limitation term, $h_N$. That is,

\begin{equation}
	g = Tc\cdot g^{max}\cdot h_E\cdot h_N/(h_E + h_N)\,.
\label{eq:g}
\end{equation}

\noindent The light limitation term is given by

\begin{equation}
	h_E = 1 - \exp(-\alpha\cdot \lambda^{max}\cdot E/g^{max}) \,, \label{eq:hE}
\end{equation}

\noindent where $\alpha$ is the initial slope of the photosynthesis
versus irradiance curve (mg $C$ mg $Chla^{-1}$ mol photon$^{-1}$
m$^2$), and  $\lambda^{max}$ is the maximum chlorophyll-a:carbon ratio
(mg $Chla$ mg $C^{-1}$). The parameter $\alpha = a_{Ch} \cdot Q$ is the product of the chlorophyll-specific absorption coefficient for phytoplankton, $a_{Ch}$ (m$^2$ mg $Chla^{-1}$), and the maximum quantum yield for photosynthesis, $Q$ (mg $C$ mol photons$^{-1}$).

The nitrogen limitation term is given by 

\begin{equation}
	h_N = N /((g^{max} \cdot Tc/a_N)+N)\,,
\end{equation}

\noindent where $a_N$ is the maximum specific affinity  for nitrogen uptake (mg $N^{-1}$ m$^3$ d$^{-1}$). 

The phytoplankton nitrogen:carbon ratio, $\chi$, predicted  by the model is given by:

\begin{equation}
	\chi = \frac{\chi^{min}\cdot h_E + \chi^{max}\cdot h_N} {h_E + h_N}\,, \label{eq:chi}
\end{equation}

\noindent where $\chi^{min}$ and $\chi^{max}$ are the  minimum and maximum nitrogen:carbon ratios (mg $N$ mg $C^{-1}$). 
 
The model predicts the phytoplankton chlorophyll-a:carbon  ratio $\lambda$, and this can be combined with the nitrogen:carbon ratio $\chi$ to convert  phytoplankton biomass $P$ (mg $N$ m$^{-3}$) to a  predicted $Chla$ concentration as:

\begin{equation}
Chla = P\cdot (\lambda^{max}/\chi^{max})\cdot h_N \cdot Tc /(R_N \cdot h_E + h_N),
\label{eq:ChlP}
\end{equation}

\noindent where $R_N$ = $\chi^{min}/\chi^{max}$. This growth model involves six parameters ($g^{max}$, $\alpha$, $\lambda^{max}$, $a_N$, $\chi^{max}$, $R_N$). The parameters $\alpha$, $\lambda^{max}$ and $\chi^{max}$ appear only in terms of the ratios $\alpha$ / $\lambda^{max}$, and $\lambda^{max}$ / $\chi^{max}$, but since $\chi^{max}$ is fixed based on the Redfield ratio, this does not result in redundant parameters in our inference procedure.
  
While this completes the specification of the local reactions $\bf{R}$
given in (\ref{eq:partialc}), in the simple one-box, mixed-layer
(i.e., 0-D) model adopted here, we do need to allow for effects of physical exchanges between the mixed layer and the underlying water mass. These exchanges add additional source-sink terms to the right-hand sides of Equations (\ref{eq:dPdt})-(\ref{eq:dNdt}), and these are specified in Appendix \ref{app:sourcesink}.
  
\subsubsection{From a Deterministic to a Stochastic BGC Process Model}\label{sec:SBGC}

The BHM framework encourages  us to formulate the state or process model  in probabilistic or stochastic terms, in order  to capture the effects of approximations  and errors in the process representation.  Note that  a stochastic-model formulation is not equivalent to  recognising prior uncertainty in the (constant)  parameters in a deterministic model.  A  deterministic model effectively asserts that,  given the initial state and the parameters,  the future state can be predicted exactly at  all future times.  A stochastic model asserts  that, given the model state and parameters at  the current time, we can make statements only  about the probability distribution of the state  at future times. 
 
A deterministic model of the kind described  in Section
\ref{sec:procmodel} can be converted to a stochastic model  in a
number of ways.  The simplest approach is  to introduce an additive
error term on the right-hand  side of equations , either as a
continuous Wiener process for the differential equations (\ref{eq:dPdt})-(\ref{eq:dNdt}), or as a
Gaussian error term at each time step in the discretised version.  We
have not adopted that  approach here; we have tried instead to
introduce  randomness into the process model in a way that  better
reflects the approximations we make in  formulating such
models, and that preserves mass balance. Specifically, we replace the constant ecophysiological
parameters in the deterministic model with stochastic processes
that change as the underlying plankton community composition
changes. In the remainder of this section we provide motivation for,
and a detailed explanation of, this approach.
 
A key approximation made in formulating models like the one given in
Section \ref{sec:procmodel}, involves biological aggregation.
Phytoplankton and zooplankton communities, which  consist of many
different species, are each  represented in the model by a single
compartment.   More complicated models may divide phytoplankton  or
zooplankton biomass into two or more functional groups with different
ecological roles, but each group still constitutes an aggregation of
diverse species. The model formulations used in Section \ref{sec:procmodel} are largely derived from many, many laboratory studies  of individual species or isolated samples, which give us reason for  confidence in the structural form of the  models. However, these studies also show very  large levels of variation in many of these  eco-physiological parameters, across individual  species, or across field samples.  Hence, the properties assigned to functional groups in these models must be thought of as representing some kind of average across the community of species making up the functional group.

The key point here is not just that variation  exists, and so there is uncertainty in specifying  these community properties, but that community  composition varies over time, and so the  community parameters must also be expected  to vary over time.  In models like those given in Section \ref{sec:procmodel}, we do not attempt to explain or predict these changes in  community composition (and consequently in  community properties) mechanistically, but we can account for them by treating them as stochastic  processes.  Now, we expect some level of persistence  in community composition, so it does not seem  realistic to treat community properties as  being drawn independently from some underlying  distribution at each time step. Instead, we allow for  community persistence by treating community properties as the outcome of a first-order autoregressive  stochastic process. 

This means that if $b$ is a generic biogeochemical \textit{parameter}  in the deterministic model, we replace $b$ by a  stochastic BGC process $B$ in the  model, with

\begin{equation} 
B(t + \Delta t) = B(t)\cdot (1 - \Delta t/ \tau ) + \zeta_B(t) \cdot
\Delta t/\tau, \quad \mbox{ for } | 1 - \Delta t/ \tau | < 1.
\label{eq:Btdt} 
\end{equation}

\noindent Here, $\Delta t$ is the discrete time step (assumed  to be 1
day in our example), $\tau$ is the characteristic time  of the
autoregressive process (that is, the time  scale on which community
composition changes),  and $\{\zeta_B(t)\}$ represents a sequence of
independent and identically distributed random variables with
distribution [$\zeta_B$]. Detailed properties of this process required
for our study are provided in Appendix \ref{app:ATLG}.

We can obtain prior information on the distribution  [$\zeta_B$] by
considering past laboratory and field  studies.   In fact,
meta-analyses  of past studies for many ecophysiological parameters
have been conducted  by researchers looking to establish systematic
relationships between these parameters  and individual size.  These
analyses  show that parameters typically vary over orders of magnitude, so there is good
reason  to propose log-normal distributions for [$\zeta_B$]  (i.e.,
normal distributions for log($\zeta_B(t)$)),  for most parameters.

There are some further complications we need to  consider in making
the step from a meta-analysis  of laboratory studies to specifying a
prior for  distributions like $[\zeta_B]$.  The meta-analyses
summarize results of measurements on individual species drawn from a
wide variety of locations, but the processes $B$ refer to means over
the community of species present at a
particular location. We would expect the variance of the community
mean to be less than the variance over the constituent species; this
effect is dealt with explicitly in Appendix \ref{app:ATLG}.
 It is also possible that the species comprising a functional group at
 a particular location will be less diverse, and may exhibit lower
 variance, than the species represented in meta-analyses. We denote
 the ratio of the coefficient of variation (CV) of community mean
 parameters to the CV of species parameters by $PDF$ for phytoplankton, 
and $ZDF$ for zooplankton. In Appendix \ref{app:ATLG}, we relate these ratios to
measures of community diversity. 

Because of the lognormal nature of the autoregressive error
$\zeta_B(t)$ in (\ref{eq:Btdt}), we consider the mean of $B$, $E(B)$,
and the coefficient of variation of $B$, namely
$CV(B)\equiv\sqrt{Var(B)}/E(B)$. Appendix \ref{app:ATLG} 
shows how it is possible to choose the mean and variance of
log$\zeta_B(t)$ such that $E(B)$ and $CV(B)$ are consistent with the
mean and variance of individual species properties, given the values
of $PDF$ and $ZDF$. We treat $PDF$, $ZDF$ and the expected value
$E(\mathbf{B}) \equiv \boldsymbol{\mu}_{\mathbf{B}}$, where $\mathbf{B}$ is the set of all BGC
autoregressive processes, as parameters in
$\boldsymbol{\theta_W}$. We also assume characteristic time scales for
changes in phytoplankton community composition ($\tau_P$), and
likewise for zooplankton community composition ($\tau_Z$).

We need to establish priors for the parameters controlling the behaviour of
the autoregressive processes: $PDF$, $ZDF$ and
$\boldsymbol{\mu}_{\mathbf{B}}$. We set broad, relatively uninformative priors
for $PDF$ and $ZDF$. We also set relatively uninformative priors for the
components of $\boldsymbol{\mu}_{\mathbf{B}}$, by assigning them the same
distribution (mean and variance) used to describe the individual species
parameters, based on the meta--data (Appendix \ref{app:ATLG}). This means that
the prior distribution allows the community parameter to take on the most
extreme values revealed by individual species. For further information on
priors and their derivations, see Appendix \ref{app:prior_model} and Section
\ref{sec:prior}.

We can now translate the stochastic BGC process model into the BHM
formalism presented in Section 2. The process \textbf{W}, as defined
in Section  \ref{sec:BHM}, can be split into the state vector
\textbf{X} and a vector \textbf{B} that recall is the set of autoregressive BGC processes.  That is,
\begin{equation}
[{\bf{W}}]=[{\bf{X}},{\bf{B}}]\,, \label{eq:WXB}
\end{equation}

\noindent where the state is $\textbf{X} = \{N,P,Z,D\}$ and the (random) BGC
processes are $\textbf{B} = \{g^{max}$, $\lambda^{max}$, $R_n$, $a_N$, $I_Z$,
$Cl_Z$, $E_Z$, $r_D$, $m_Q\}$. Similarly, $\boldsymbol{\theta_W}$ in
(\ref{eq:YWtheta3}) can be split into two parameter sets, those appearing
explicitly in the equations updating $\bf{X}$, namely $\boldsymbol{\theta_X} =
\{K_W,a_{Ch},s_D,f_D\}$, and those appearing in the autoregressive equations
for the BGC processes $\bf{B}$, namely $\boldsymbol{\theta_B} = \{PDF$, $ZDF$,
$\mu_{g^{max}}$, $\mu_{\lambda^{max}}$, $ \mu_{R_N}$, $\mu_{a_N}$,
$\mu_{I_Z}$, $\mu_{Cl_Z}$, $\mu_{E_Z}$, $\mu_{r_D}$, $\mu_{m_Q}\}$, taking
note that $PDF$ and $ZDF$ effectively scale the coefficient of variation,
$CV(B)$, given in \ref{tab:prior} of Appendix \ref{app:prior_model} .  The
state-space representation is now as given in Figure \ref{fig:statespacerep} below.

\begin{figure}[htp]
\centering
\includegraphics[scale=1.4]{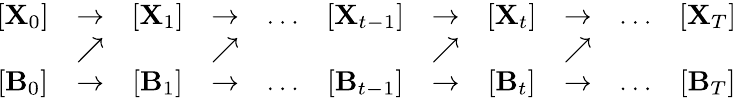}
\caption{Evolution of the state (\textbf{X}) and the BGC (\textbf{B})
  processes. Recall that \textbf{W} = [\textbf{X,B}] and that Figure 1
  shows how observations \textbf{Y} are related to the process
  \textbf{W}.}
\label{fig:statespacerep}
\end{figure}


In terms of conditional probabilities, the formulation developed in this section means that:
\begin{equation}\label{eq:XtBt}
[{\bf W}_t | {\bf W}_{t-1}, \boldsymbol{\theta}_W] = [{\bf X}_t,{\bf
    B}_{t}|{\bf X}_{t-1},{\bf B}_{t-1},{\boldsymbol{\theta}_W}]= [{\bf
    X}_t|{\bf X}_{t-1},{\bf B}_{t-1}, \boldsymbol{\theta}_X][{\bf B}_t|{\bf B}_{t-1},{\boldsymbol{\theta}_B}]\,,
\end{equation}
\noindent where the last equality expresses the fundamental evolution of the process model (Section \ref{sec:SBGC}).

\subsection{The parameter (prior) model}\label{sec:prior}

The priors assigned to the parameters specified in this study were drawn from
a meta-analysis of the literature.  A summary of the prior information
available for the BGC parameters and processes, and the sources of this
information, is given in Appendix \ref{app:prior_model}.  Each component of
the prior is assumed independent of the other components, and no attempt has
been made to introduce any dependence structure between the parameters.

\subsection{The data model}\label{sec:datamodel}

\indent The data model explicitly links the  process model with the
observations.  The parameters ${\boldsymbol{\theta}}_{\mathbf{Y}}$ in
(\ref{eq:YWtheta2}) control the observation process, and we
consider two broad classes of observation error. 

\begin{enumerate}
	\item Analytical measurement errors should reflect the precision of \textit{in situ} instruments or laboratory analyses.  For example,  laboratory determinations of chlorophyll-a pigment concentration might be expected to have a precision of a few percent.
	\item Representation errors can arise from (i) mismatches in scale (we may model a  large volume of ocean, many kilometres across, but make measurements on bottle samples comprising a few litres), and (ii) mismatches in type (we may predict zooplankton concentration in the currency of nitrogen, but measure volume or wet weight of biomass). 
\end{enumerate}

In most real-world situations, errors associated with mismatches in scale and type outweigh analytical measurement errors. The use of a simple 1-box mixed layer model here introduces an additional ambiguity. We are neglecting horizontal advection, which might be thought of as an additional process-model error. The significance of horizontal advection compared with local processes depends on the area of ocean represented by the box. If we regard the box as representing an ocean area several hundred kilometres in extent, we might hope that the errors involved in neglecting advection are small. But we must then expand the observation error to account for the spatial variability observed on these length scales.

In Section~\ref{sec:learning}, the data model for our application to data from Ocean Station Papa is given by:
\begin{equation}
{[\bf{Y} | \bf{W}, \boldsymbol{\theta_Y}]}=\prod_{t=1}^{T}[{\mathbf{Y}}_t|{\mathbf{X}}_t,\boldsymbol{\theta}_Y]\,.
\label{eq:YtXt}
\end{equation}

Treatment of $\boldsymbol{\theta_{\bf{Y}}}$ for our case study is
discussed in section \ref{sec:learning}. Recall  that $\bf{W}$ is made
up of $\bf{X}$ and $\bf{B}$; note that if we had direct observations
of the ecophysiological properties represented in $\bf{B}$, these
could be incorporated into the data model.

\section{Learning and Predictability Given Observations}\label{sec:learning}

\indent We demonstrate the application of the BHM  framework to a marine
BGC model using the historical Ocean Station Papa (OSP) dataset as a case study. This site was chosen over alternative subtropical time series sites because the simple mixed layer model is believed to be a better approximation at OSP. Two  experiments were conducted:
\begin{enumerate}
	\item A twin experiment was run using climatological forcing at OSP,
          with synthetic observations of all state variables assimilated
          daily. The synthetic observations were generated by adding noise to
          a known ``true''  trajectory through the state space.
	\item A subset of the historic OSP dataset comprising observations of
          chlorophyll-a ($Chla$) and nitrate ($N$) was assimilated for the period January 1971 -
          November 1974. This corresponds to part of a sustained observing
          campaign, and we found that the marginal posteriors for parameters did not
          change greatly if additional years were included.
\end{enumerate}

\subsection{Ocean Station Papa Site Description} 
\begin{figure}[tp]
	\centering
        \includegraphics[width=1.00\textwidth]{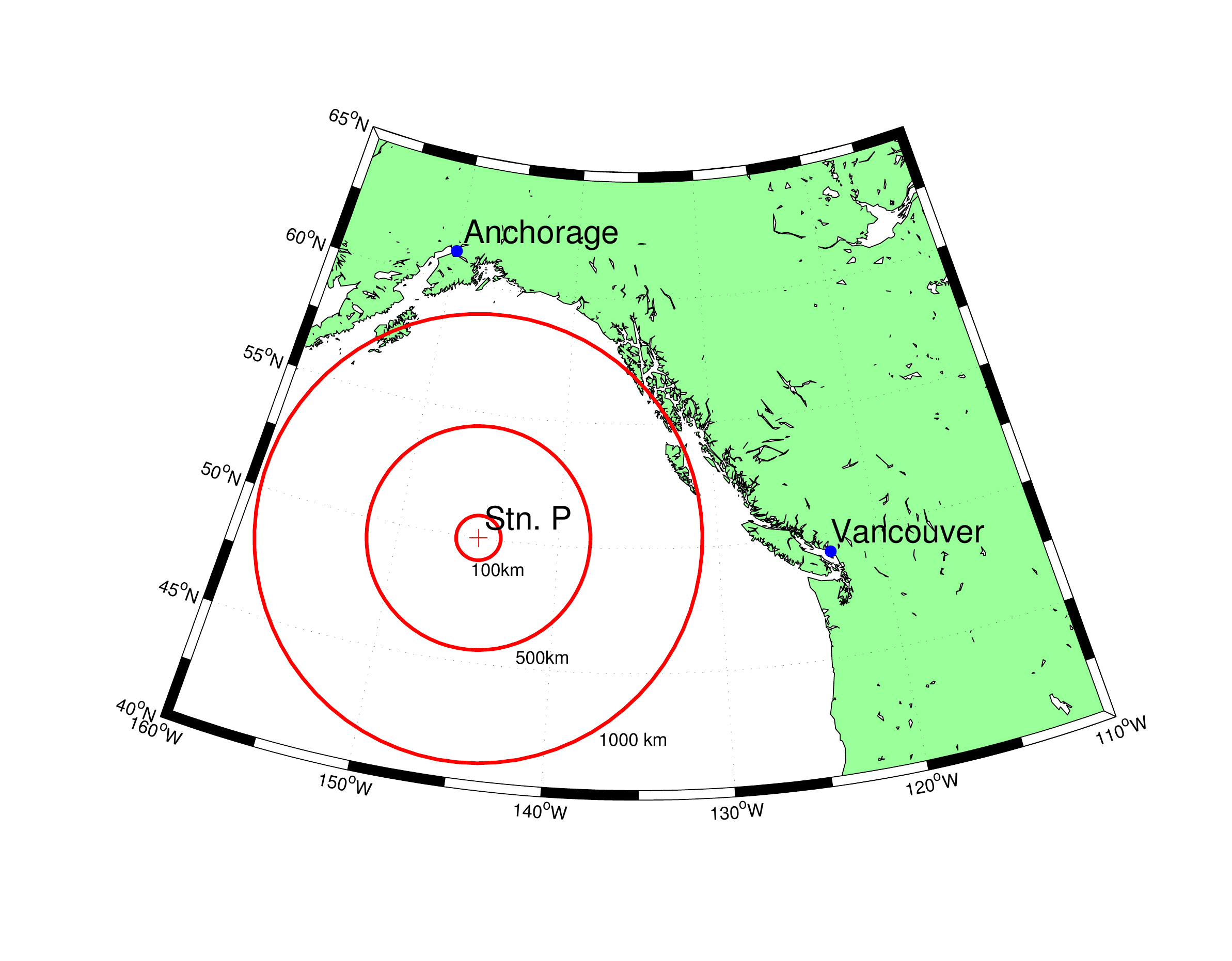}
	\caption{A map of the north-east Pacific Ocean displaying the location of Ocean Station Papa (Stn. P) with range circles at 100, 500 and 1000 km.}
	\label{fig:OSPmap}
\end{figure}

Ocean Station Papa (OSP) is
located at 50$^\circ$N, 145$^\circ$W (Figure \ref{fig:OSPmap}), in 1500 m of
water in the sub-arctic region of the north east Pacific Ocean. It experiences
a strong seasonal cycle in temperature, wind stress, and incident solar
radiation \citep{Whitney1999}. During winter and spring, a mixed layer of
depth 80 - 120 m is sustained by a high wind stress with the low incident
solar radiation unable to induce any persistent stratification of the water
column.  During summer, the thermocline shallows in response to increased
surface heating and a reduction in the wind stress. Consequently, a relatively
shallow mixed layer is maintained of typical depth 25 - 40 m.

It has been noted that there are persistently high macro-nutrient
concentrations in the mixed layer and the phytoplankton biomass is typically
low. This phenomenon is observed throughout much of the open sub-arctic
Pacific ocean. While the concentration of dissolved inorganic nitrogen (DIN)
is lower in summer than in winter, it is rarely if ever depleted to levels
that may cause nutrient limitation in primary producers
\citep{Harrison2002}. There is no discernible seasonal cycle in
chlorophyll-a. Previous modelling studies of \citet{Matear1995},
\citet{Denman1999} and \citet{Denman2003} discuss the likely controls on
phytoplankton biomass and the seasonal variation in primary productivity and
zooplankton biomass.

\subsection{Learning from Observations: Twin Experiment with Climatological Forcing}\label{sec:TE_clim}

\indent Twin experiments in a setting like that of OSP have been conducted to
compare samples from the posterior, $[{\bf W ,\boldsymbol{\theta}}|{\bf
    {\bf{Y}}}]$, produced by Bayesian inference, with known ``true'' values of
the state and parameters. The term ``twin'', borrowed from the
data-assimilation literature, refers to experiments where the model used for
inference, and the model from which synthetic observations are generated, are
the same. Model forcing and boundary conditions are taken from
\citet{Matear1995} and are climatological in nature; details are given in
Appendix \ref{app:CS_SUP}.

\subsubsection{Twin Experiment: Design}\label{sec:ExpDesign}

To generate the synthetic observations, we select a parameter set
$\boldsymbol{\theta}^*$ (the ``true'' parameters) and take a single
realisation of the stochastic model $\{\mathbf{W}^*_t : t = 0,1,\ldots,T\}$ to
produce the trajectory $\{\mathbf{X}^*_t: t = 0,1,\ldots,T\}$ through state
space (again referred to as the truth).  We have chosen a set of ``true''
parameters in the twin experiment that are shifted away from the prior means
(to provide a clearer test of the inference procedure), but that nevertheless
yield state-variable trajectories qualitatively consistent with OSP
observations (e.g., high-nutrient low-chlorophyll (HNLC) conditions). The
(synthetic) observations $\mathbf{Y}$ are generated by:
\begin{equation}
	{\bf{Y}}_t = {\bf{X}}^*_t \exp(\xi_t);\;t = 0,1,\ldots,T\,,
\end{equation}
where $\xi_t$ are independent and identically distributed (IID) as the normal
distribution $N(0, \sigma^2_{obs})$. The standard deviation, $\sigma_{obs}$,
was $0.1$ for DIN observations and $0.2$ for observations of the remaining
state variables.  The log-normal error model was adopted because errors in the
estimates of plankton density are typically better represented by log-normal
multiplicative error than by additive normal error \citep{Campbell1995}, and
the log-normal multiplicative-error model delivers synthetic observations that
are non-negative. The observation errors are assumed to be independent over
time, reflecting either analytical error or (more likely) uncorrelated
small-scale variation in concentrations.

\subsubsection{Twin Experiment: Results}

\begin{figure}[tp]
	\centering
		\includegraphics[width=1.0\textwidth]{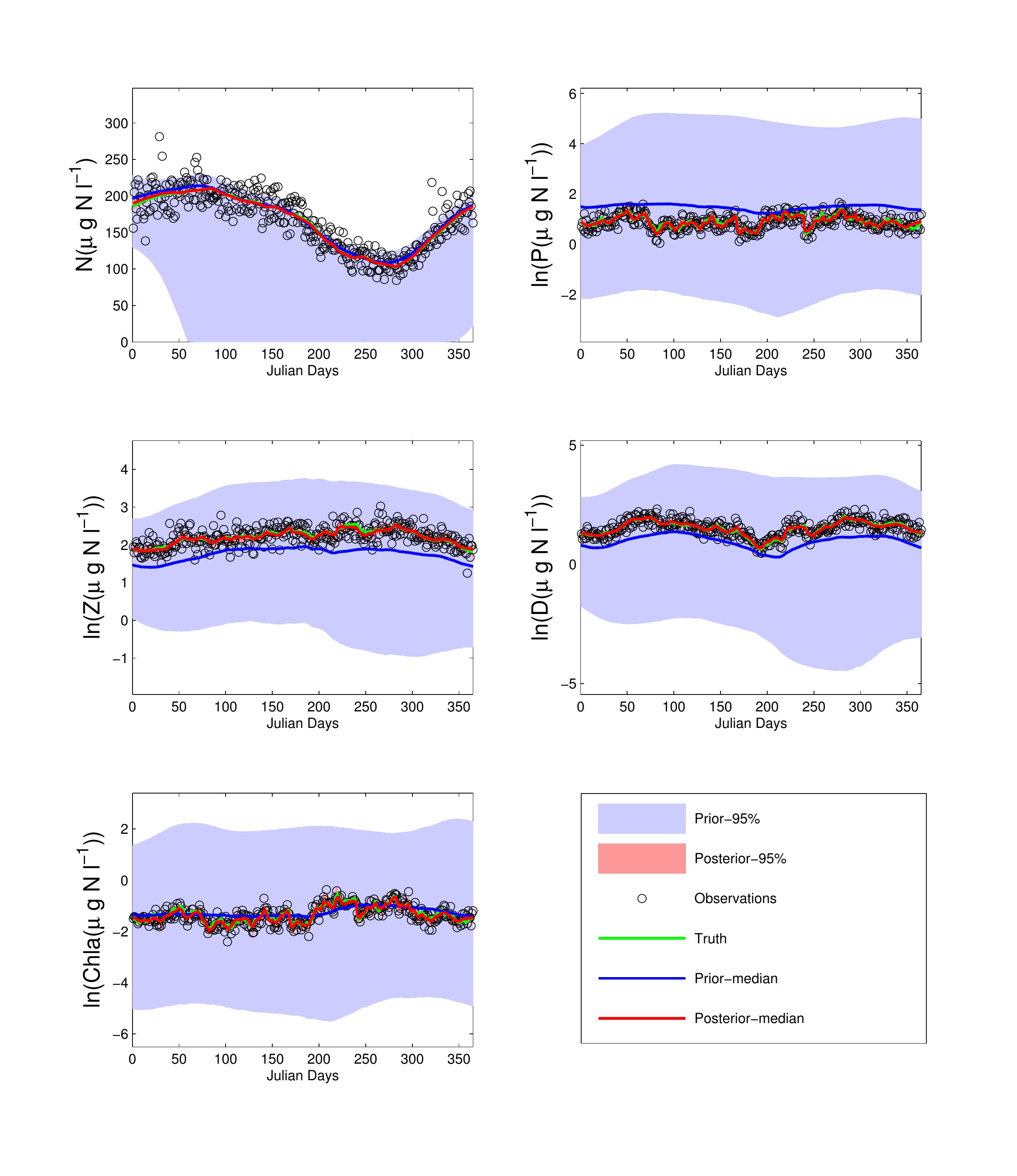}
	\caption{Twin experiment: a time series of the prior and
          posterior distributional properties of the state
          (\textbf{X}). Note that the posterior credibility intervals
          remain so close to the posterior median that they are
          difficult to distinguish. The figure legend is shown on the lower right. }
	\label{fig:TE_state}
\end{figure}

We first generate an ensemble of model trajectories by sampling from the prior
distribution for parameters and running the stochastic model forward through
the period January 1971 - November 1974, without assimilating any
observations. This so-called free-run process-model ensemble is precisely a
sample from the prior distribution over the state (Figure \ref{fig:TE_state},
blue shading), which expresses the uncertainty in the state based only on the
prior knowledge of the parameters gained from a meta-analysis of the
literature. In spite of the large prior uncertainty in some of the process-model parameters, the median values of the (marginal) prior distributions over
state variables show surprisingly similar qualitative behavior to the observed
climatology at OSP (Figure \ref{fig:TE_state}, dark blue line). The median DIN
values remain elevated, and median chlorophyll-a values remain low. However,
the 95\% contours of the prior ensemble include unrealistic behaviours not
observed at OSP, involving near-complete depletion of DIN and intense
phytoplankton blooms.

When the synthetic observations described in Section \ref{sec:ExpDesign} are
assimilated, using the methodology described in Section \ref{sec:BHMapp}, the
95\% credibility intervals for the posterior distribution of the state are
very tightly constrained about the true trajectory (Figure \ref{fig:TE_state},
red shading), compared with the prior intervals and with the
observations. Despite the 20\% observation error, the dynamical BHM
implemented through the PMCMC described in Appendix \ref{app:PMMH},
 accurately tracks the true state (Figure \ref{fig:TE_state}, green
 line).

The case for N deserves additional explanation. The seasonally varying
N concentration, prescribed below the mixed layer as a boundary
condition, imposes a sharp upper limit to the predicted mixed-layer N
concentrations. Provided grazing control keeps phytoplankton biomass
and N utilization small, the predicted concentration is very close to
this upper bound.  In most prior trajectories, grazing control is
effective, so the prior median is close to the upper limit. Some prior
parameter combinations allow phytoplankton blooms and N depletion,
resulting in the drawdown of N to near-zero levels seen in the prior
lower  95th percentile for N. The truth is chosen to be OSP-like, and
so produces N concentrations close to the upper bound. Since we add
noise to the truth, a significant fraction of the observations lie above
the upper bound.

\begin{figure}[tp]
	\centering
		\includegraphics[width=1.0\textwidth]{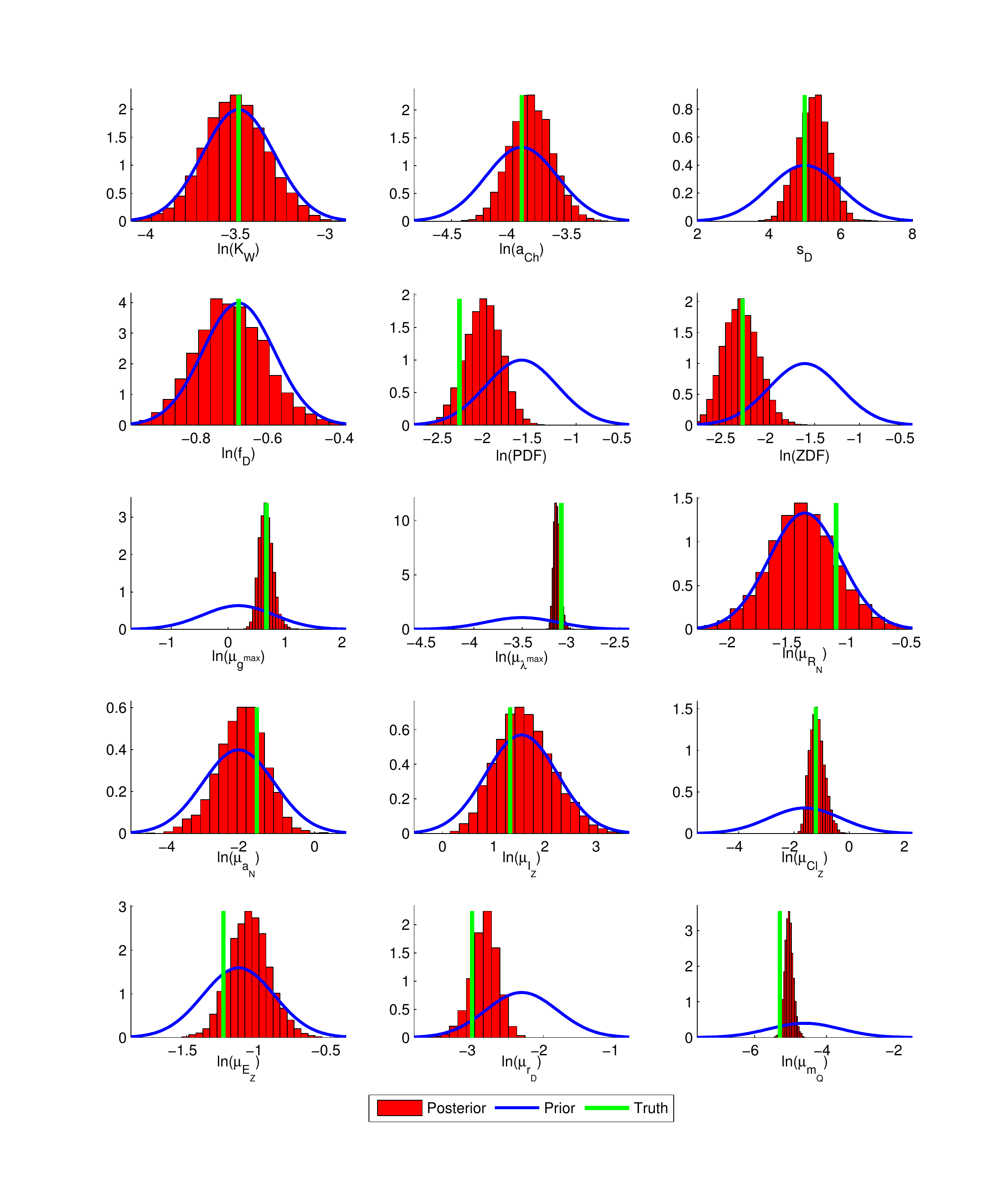}
	\caption{The prior (blue curve) and posteriors (red histogram) for each of the parameters. The true value is given by the vertical green line.}
	\label{fig:TE_parameters}
\end{figure}

The prior distributions over the parameters given in Table \ref{tab:prior} 
of Appendix \ref{app:prior_model} are
the blue curves in Figure \ref{fig:TE_parameters}. These priors are discussed
in Section \ref{sec:prior} and are considered ``global'' in that they
represent experimental results encompassing a wide range of species and
domains. For some model parameters ($a_{Ch}$,
$s_D$, $PDF$, $ZDF$, $\mu_{g^{max}}$, $\mu_{\lambda^{max}}$, $\mu_{Cl_Z}$,
$\mu_{E_Z}$, $\mu_{r_D}$, and $\mu_{m_Q}$), the marginal posteriors in Figure
\ref{fig:TE_parameters} show evidence of learning in that the posterior mode
has moved towards the truth and the posterior variances have contracted
compared with the prior. However, for others, the inference procedure appears
to extract little or no information from the data, and the marginal posteriors
appear to merely recover the prior distributions. This is true for the
parameters controlling light attenuation due to water ($K_W$), the fraction of
zooplankton waste diverted to detritus ($f_D$), the parameters related to
nitrogen uptake and nitrogen:carbon ratios ($a_N$ and $R_N$), and the maximum
zooplankton ingestion rate ($I_Z$). In the case of $a_N$, the posterior
variance is slightly reduced, but the posterior median remains centred at the
prior mean.

\begin{figure}[tp]
	\centering
		\includegraphics[width=1.0\textwidth]{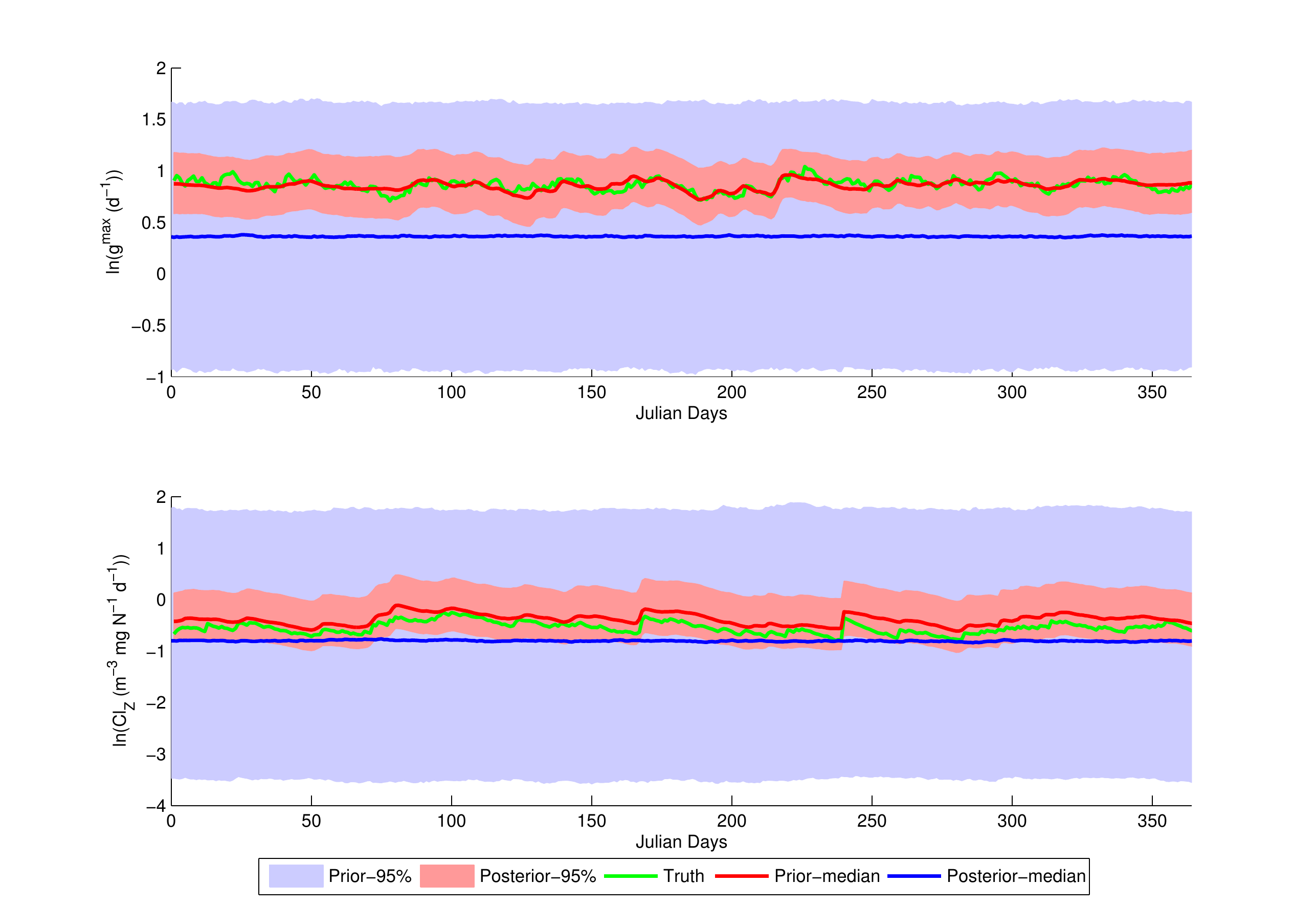}
	\caption{Twin experiment: a time series of the prior and
          posterior distributional properties of the autoregressive processes ($\mathbf{B}(t)$).}
	\label{fig:TE_AR_Gmax_ClZ}
\end{figure}

The inference procedure generates posterior distributions for time series of
the autoregressive processes $\mathbf{B}(t)$, and could provide information
about changes over time in the ecophysiological properties they
represent. However, the results from this twin experiment are only mildly
encouraging in this regard. In cases where the observations are uninformative
about the parameters underlying the autoregressive processes, one can hardly
expect to obtain information about the temporal variation in the processes
themselves. Indeed, in those cases, the posteriors for the stochastic-process
trajectories are the same as the priors. In two cases ($g^{max}$ and $Cl_Z$),
the posterior median trajectories appear to track the truth, although with
consistent bias in the case of $Cl_Z$ (Figure \ref{fig:TE_AR_Gmax_ClZ}).  But
for these, and all other autoregressive processes, the 95\% credibility
interval for the posterior exceeds the amplitude of the temporal variation in
the truth by some margin. The inference procedure does not allow us to
conclude that there are significant changes in these processes over time.

These results reflect the particular nature of the climatological forcing and
system behavior at OSP. Given that concentrations of dissolved inorganic
nitrogen at OSP remain well above levels expected to limit phytoplankton
growth, it is unsurprising that parameters controlling nitrogen limitation of
growth rates are poorly constrained. Similarly, phytoplankton biomass remains
at levels well below those required to saturate zooplankton grazing, and
zooplankton growth rates are controlled by the clearance rate $Cl_Z$, not by
the maximum ingestion rate.

\subsection{Learning from Observations: Ocean Station Papa Dataset}

To demonstrate the application of the BHM approach to a real dataset, we have used a subsample of historical OSP data.

\subsubsection{Ocean Station Papa - Data Model}

Observations of nitrate (DIN) and chlorophyll-a taken between January 1971 and November 1974 are used. Observation errors are large and dominated by spatial sampling errors, because we neglect horizontal advection and assume a large model domain with high levels of within-domain variability. The presence of larger observation errors means that the data will be less informative. We draw on a number of studies below for estimates of the appropriate levels of spatial variability.

The spatial and temporal variability of  Particulate Organic Carbon (POC) in this region  has been investigated at a number of scales \citep{Bishop1999}. The spatial variability  in the vertical and horizontal directions was calculated  from the beam-attenuation coefficient obtained from a transmissometer. 
\begin{itemize}
\item Small-scale horizontal variability (1-10km) of POC appears to be 5\% to
  10\%, which is   deemed negligible in comparison to  the ocean scale and temporal variability.

\item Large-scale horizontal variability (100-300km) of POC appears to
  range from  10\% to 40\%, however we attribute some of this
  variability to the passage of weather systems on time scales of 5 to
  10 days.
	
\item Ocean-basin-scale variability (800-2000km) exceeds both the large-scale
  and small-scale spatial variability, but this is due to the change from
  HNLC conditions in the deep ocean to a more
  typical temperate seasonal cycle on the continental shelf.
\end{itemize}

\citet{Bishop1999} also noted significant interannual variability that may be
linked to El Ni\~{n}o events.  Nitrate data collected along the Line P
transect (a 1425 km long transect between the coast adjacent to the Juan de Fuca strait and Ocean Station Papa \citep[e.g.][]{Pena2007}) from 1992 to 1997 display a similar pattern to the POC data.  Again,
it appears that on the scale of 100-300km around OSP, variability in total
concentrations of nitrates and nitrites appears to be 10\% to 30\%, with
inter-annual variability exceeding the large-scale spatial
variability \citep{Whitney1999}.

Taking all these sources of information into account, we have assigned a CV ($\sigma_{\text{obs}}$) of 0.5 to the observation error for both DIN and chlorophyll-a. This is a conservative (upper) estimate, representing an upper bound to spatial variation, and allowing for other non-spatial contributions, including analytical measurement error.

\subsubsection{Ocean Station Papa - Results from Hindcast}

\begin{figure}[tp]
	\centering
		\includegraphics[width=1.0\textwidth]{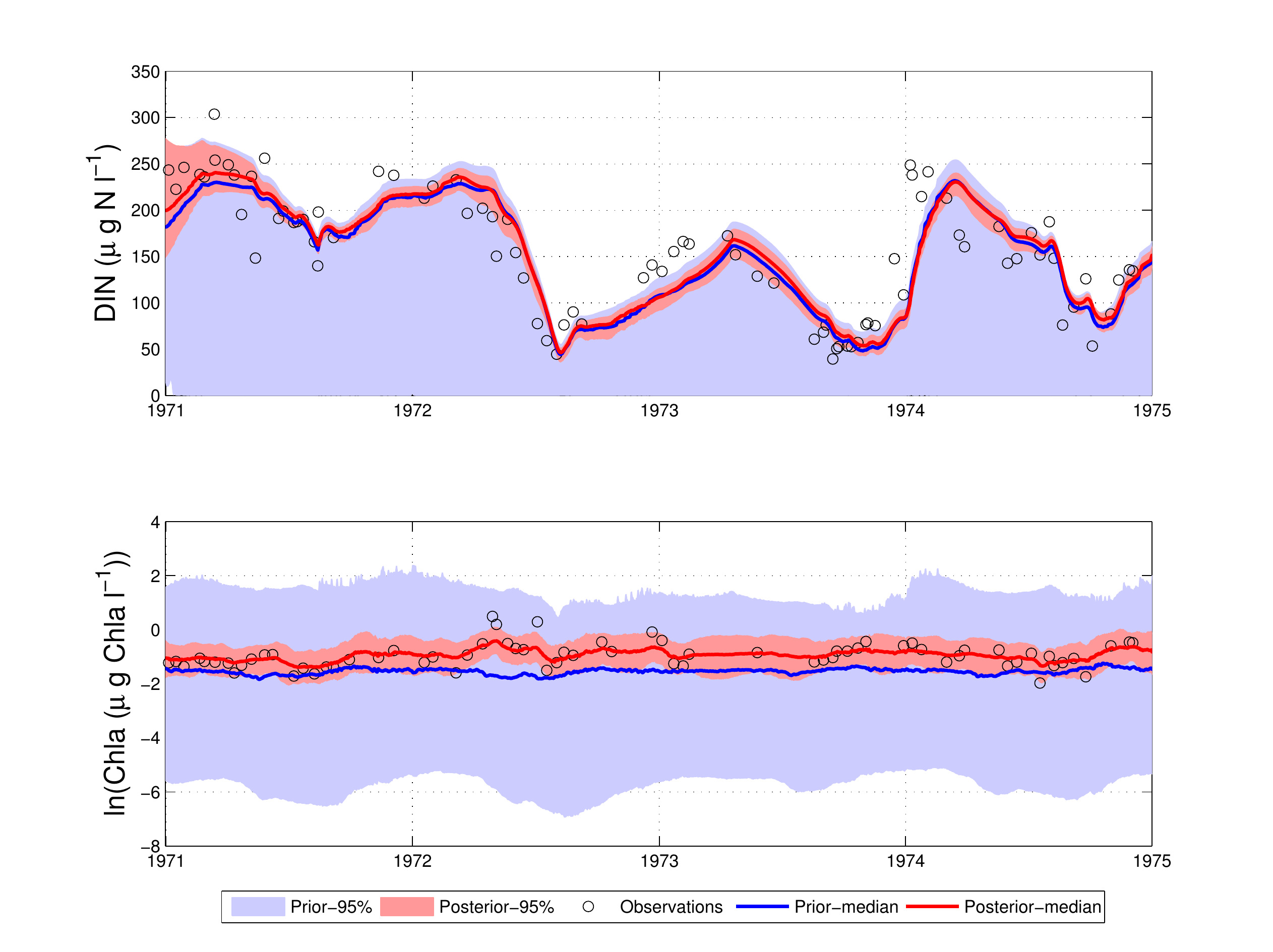}
	\caption{A time series of prior and posterior distributional
          properties of observed state variables, comparing observations (open black circles), prior (blue) and posterior (red).}
	\label{fig:OSP_ChlN}
\end{figure}

\begin{figure}[tp]
	\centering
		\includegraphics[width=1.0\textwidth]{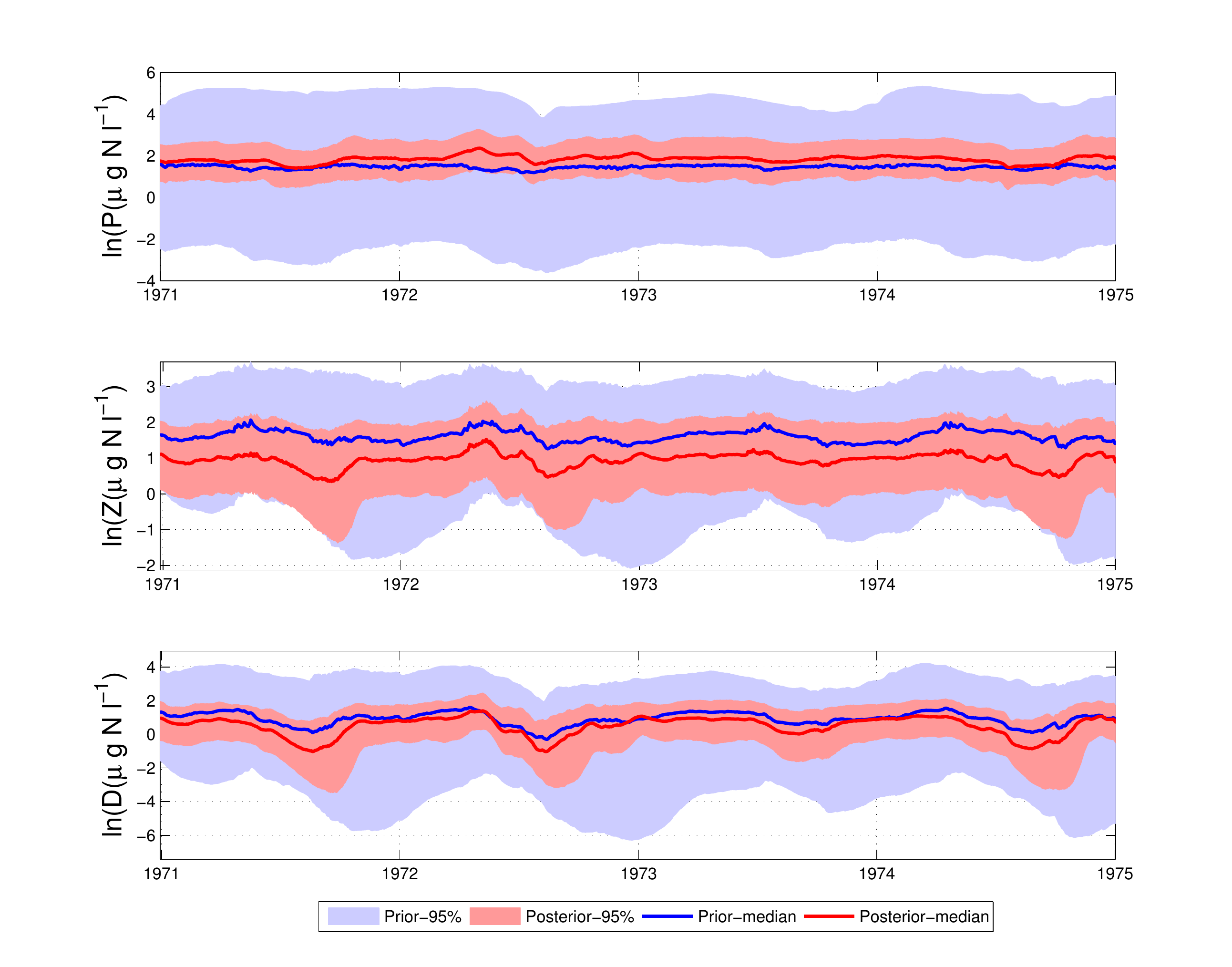}
	\caption{A time series of prior and posterior distributional
          properties of unobserved state variables comparing prior (blue) and posterior (red).}
	\label{fig:OSP_PZD}
\end{figure}

A prior ensemble over the state was constructed in a similar manner to the
twin experiment, using real forcing from January 1971 to November
1974.  Model parameters were sampled from the prior distributions described
earlier. As in the twin experiment in Section \ref{sec:TE_clim}, a wide range
of model behaviors was observed (Figures \ref{fig:OSP_ChlN} and
\ref{fig:OSP_PZD}), ranging from near-complete depletion of DIN during summer,
to year-round grazing control. As in the twin experiment, the median of the
prior over the state based on 1971-1974 forcing qualitatively agreed
with observed OSP behavior, in that DIN was never limiting, and there were no
strong phytoplankton blooms as zooplankton grazing maintained relatively
constant phytoplankton biomass \citep{Matear1995,Denman1999,Denman2003}.

When observations of chlorophyll-a and DIN are assimilated, the 95\% credibility
interval is dramatically reduced. Due to the relatively large observation
error prescribed (see Section \ref{sec:datamodel}), the transient,
low-magnitude increase in chlorophyll-a seen in the summer of 1972 is absorbed
into the observation error and not tracked in the state. While the three
individual observations of this anomalous bloom do not fall within the
posterior 95\% credibility interval, this cannot be interpreted immediately as
lack of model fit. This is because the credibility interval depicted is over
the latent chlorophyll-a \textsl{state variable}, not over the ``noisier''
\textsl{observed} chlorophyll-a; this distinction is important and is discussed
by \citet[Section 2.2.2]{Cressie2011}. Although short-lived transient features
are not tracked by the model, slow seasonal and intra-seasonal variations are
well captured. The methods described in Section \ref{sec:inference} not only
condition the state on observations from previous times, as do filtering
approaches, but also on future times. This is referred to as \emph{smoothing}
in the Bayesian filtering literature \citep{Briers2010,Fearnhead2010}. The advantage of such smoothing is evident
in time periods where there are very few observations (e.g., mid-1973).

Through the process model, Bayesian methods allow inference on the unobserved
state variables $P$, $Z$ and $D$; see Figure \ref{fig:OSP_PZD}. Notice that
there is a substantial reduction in the uncertainty expressed through the
posterior compared with that expressed through the prior, even for unobserved
state variables. For example, there is a strong seasonal cycle in the
zooplankton biomass, which has been observed in a number of studies
\citep{Harrison2002}. The peak in the zooplankton biomass occurs during mid
summer, which coincides with a peak in primary production (not shown).

\begin{figure}[tp]
	\centering
		\includegraphics[width=1.0\textwidth]{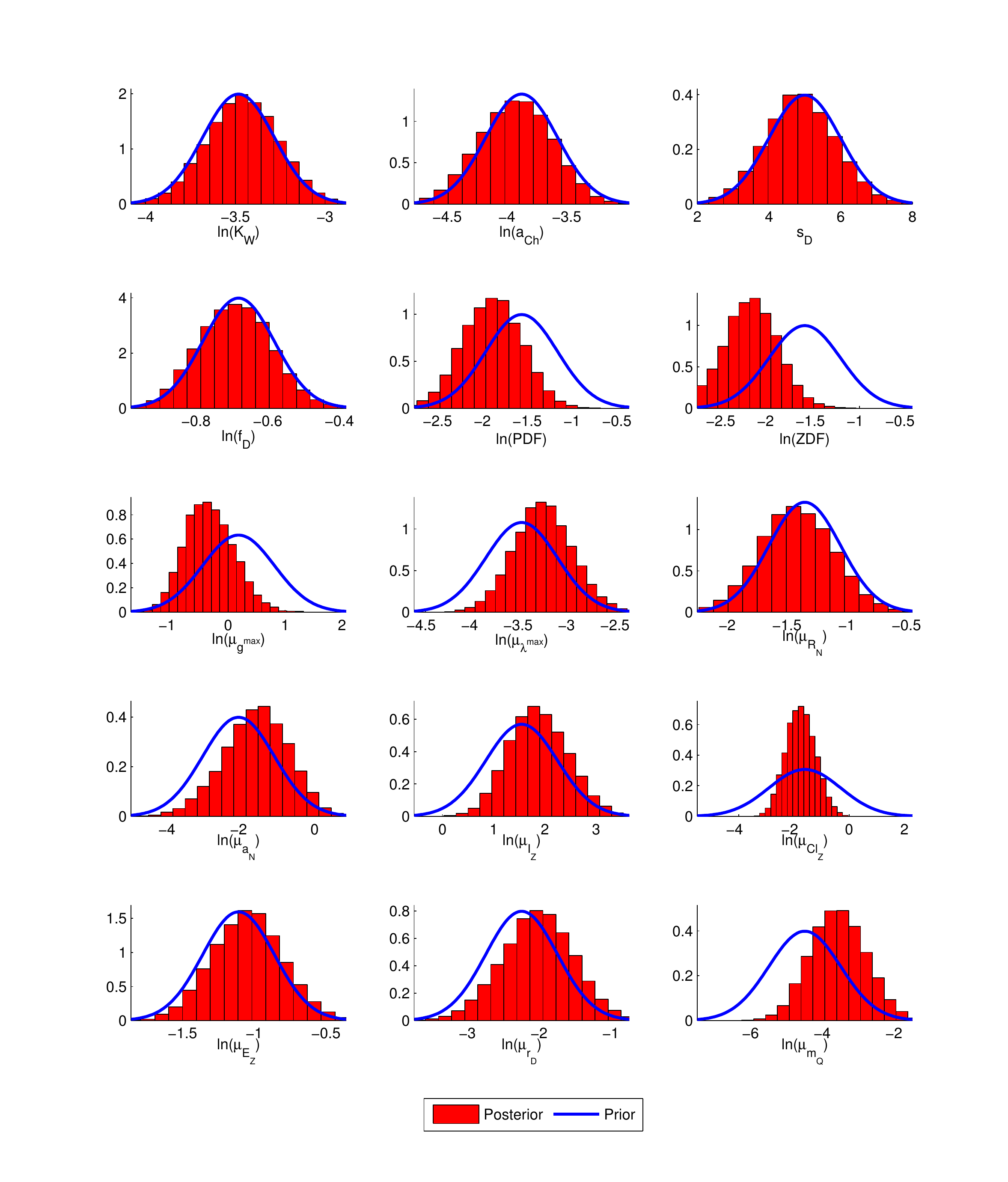}
	\caption{A comparison between the prior (blue curve) and the posterior
          (red histogram) for all parameters $\boldsymbol{\theta}_\mathbf{X}$ and $\boldsymbol{\theta}_\mathbf{B}$.}
	\label{fig:OSP_parameters}
\end{figure}

The marginal posteriors for model parameters shown in Figure \ref{fig:OSP_parameters}
demonstrate that the sparse and limited OSP observations carry very little
information about many of the parameters. This was not unexpected; previous
studies have also experienced difficulty in using the OSP dataset to estimate
parameters in deterministic models \citep{Matear1995}. The large observation
variances used here, which compensate for effects of advection, reduce the effective
information content of the data, but we believe this is realistic, given the
model structure. The posterior marginals show evidence of learning for four
parameters: $ZDF, g^{max}, \mu_{Cl_z}$ and $\mu_{I_z}$.

Perhaps unsurprisingly, given the high noise levels and sparse observations,
the OSP data do not allow us to derive useful information about temporal
variation in the autoregressive processes $\mathbf{B}(t)$. Even for those
parameters, $g^{max}$ and $Cl_z$, where the observations appear to inform the
posteriors for the underlying parameters, the posteriors for the
autoregressive trajectories show no significant variation over time (not
shown).

\begin{figure}[tp]
	\centering
		\includegraphics[width=1.0\textwidth]{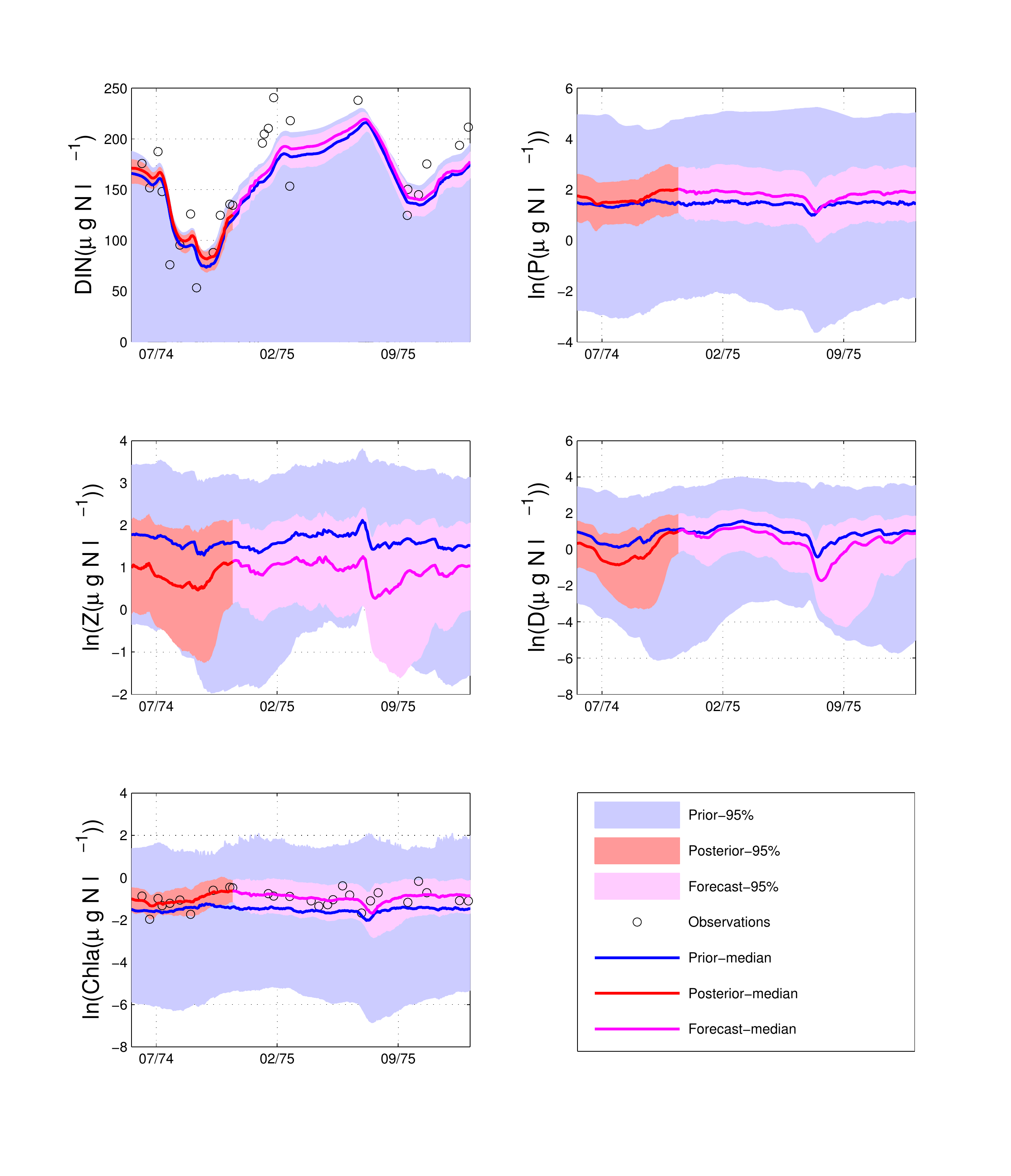}
	\caption{A forecast (magenta shading) for the model's state variables
          for the period December 1974 - December 1975. The posterior and
          prior of the state are given by the red and blue shading,
          respectively. Observations are denoted by the open black circles;
          during the forecast period, the observations were not assimilated.}
	\label{fig:OSP_state_forecast}
\end{figure}

One advantage of the BHM framework is that we can use the sample generated
from the joint posterior of the state and parameters, conditioned on past
observations, to assess the uncertainty in model forecasts and scenarios. In
this case, we have used the posterior conditional on observations from January
1971 to November 1974 to make a probabilistic forecast for 1975. We do this
simply by propagating all posterior trajectories forward the additional year,
using the boundary and forcing fields for that year. The results from this
forecast ensemble (median and 95\% credibility intervals) are shown in Figure
\ref{fig:OSP_state_forecast}. Agreement with the (non-assimilated)
observations in the forecast period is very good.

\section{Discussion and Conclusions}\label{sec:DandC}

A key consideration in building BHMs is the treatment of model
error. In our study we used the fact that the aggregation of
communities of species into single trophic levels or functional
groups, and the  replacement of well defined eco-physiological
parameters  for individual species by community-average parameters, is
an important source of model error. Consequently, we have replaced
the constant community parameters used in most biogeochemical models
by stochastic autoregressive processes that vary slowly in time. This
is in contrast to the common approach of simply adding white noise to
the rate equations.

A potential drawback of this approach is that it increases the complexity and
dimensionality of the model and the inference problem. We have augmented the
four-dimensional primary state space $(N,P,Z,D)$ with nine additional state
variables ($\mathbf{B}$). Instead of estimating nine constant parameters, we
must estimate nine means and nine variances controlling the evolution of the
stochastic processes $\mathbf{B}(t)$.  We have mitigated this problem by using
prior information to set the relative magnitudes of the variances of
phytoplankton and zooplankton community parameters, and using stochastic
factors related to community diversity to set the absolute magnitude. One
advantage of the process model, as formulated, is that it allows a strong and
direct connection to literature meta-data on the distributions of
eco-physiological parameters across species. This allows us to set informative
objective priors for most of the parameters, exploiting a key advantage of
Bayesian approaches, and partially counterbalancing the increase in unknowns.

The inference procedure was designed to derive joint posteriors for
system parameters and the (augmented) system state. Many examples of
data assimilation in dynamical models concentrate on either state
estimation or parameter estimation. Joint inference is particularly
difficult in nonlinear models with sparse data, and it has typically
required strong simplifying approximations, such as the replacement
of nonlinear dynamics by approximating linear models. The underlying
deterministic NPZD model is highly nonlinear, displaying two
qualitatively different modes of behaviour or local stability
domains, and the observed behaviour at OSP correspond to only one of
these domains. The new particle MCMC techniques employed here
are able to cope with this nonlinear, threshold behaviour, but are
computationally expensive. 

Given these challenges, the results of the OSP case study offer a
number of grounds for encouragement. First, the stochastic process
model allows the construction of priors over the model state, by
drawing random samples from the prior distribution for model
parameters and initial conditions, and running ensembles of model
simulations. We can think of this prior ensemble as encapsulating our
ability to predict system behaviour at OSP, given independent
scientific knowledge about BGC processes, and local
environmental forcing, but no other local knowledge. Encouragingly,
the state median in these prior distributions bears a strong
qualitative and even quantitative resemblance to OSP observations
(Figures \ref{fig:TE_state}, \ref{fig:OSP_ChlN}), even though the priors were chosen to reflect
the full range  of species attributes reported in the literature. But
the 95\% credibility intervals for the prior distribution also
include trajectories involving phytoplankton blooms and nitrate
depletion, which are incompatible with observations at OSP. 

Data assimilation into dynamical process models can serve a variety of
different diagnostic and prognostic purposes \citep[see
e.g.,][]{Gregg2008, LuoY2011}. One class of diagnostic applications targets the
hindcasting  or nowcasting of system state, given limited
observations. Despite  sparse observations with large sampling errors
on one  state variable ($N$) and one diagnostic variable ($Chla$), the
Bayesian inference  procedure recovers quite tight posteriors for these
observed  variables (Figure \ref{fig:OSP_ChlN}). The Bayesian inference procedure is also able to
transfer information from observed to unobserved state variables,
reducing  the uncertainty in the unobserved state variables ($P$, $Z$ and
$D$) by about half (Figure \ref{fig:OSP_PZD}).

A second class of diagnostic applications focuses on learning about,
and interpretation  of, model parameters. Here, the parameters
describe the ecological  characteristics of the plankton communities
present at OSP. To  the extent that these parameters have smaller
variances a posteriori,  we can conclude that the observations have
provided information  about the parameters and the communities they
represent. The  results for OSP are informative and cautionary. In the
twin  experiment, with observations on all state variables, the
posteriors  for some parameters are essentially identical to the
priors,  so provide no additional information
(Figure \ref{fig:TE_parameters}). These results
can be explained in terms of model dynamics. Since nutrients under OSP
conditions are always saturating, and phytoplankton concentrations
remain  low, the parameters affecting phytoplankton growth at
low  nutrient concentrations, and zooplankton ingestion at high
phytoplankton  concentrations, have negligible effect on model
predictions,  and are not identifiable. This pattern of
identifiability is  an intrinsic characteristic of the environmental
forcing  and dynamics at OSP. In other ocean conditions, such as
oligotrophic  mixed layers where nutrient concentrations are always
low and  limiting, one would expect different sets of model parameters
to  be identifiable.

Using the limited set of historical observations available for OSP,
the inference  procedure is able to extract information about a few
key  parameters only (Figure \ref{fig:OSP_parameters}). For the most
part, these parameters
directly  control the key processes involved in zooplankton grazing
control  of phytoplankton. The posterior distributions for the
parameters  controlling the variance in model parameters (PDF and ZDF)
are shifted towards lower values, compared with the priors. At the
inferred  lower levels of stochastic noise, trajectories are less
likely to  escape the local stability domain corresponding to grazing
control. 

The Bayesian inference procedure provides posterior
distributions  for the trajectories of the stochastic BGC
processes,  $\mathbf{B}(t)$. Reliable information on changes in these
processes  would be of particular interest to plankton
ecologists. However,  even in a twin experiment with daily data on all
state  variables, we were only able to obtain suggestive (but not
confirmatory) information about
temporal  variation in two parameters. This limited success is
understandable,  given that we are effectively trying to extract
information  about changes in unobserved variables on relatively short
time  scales, when the evidence of these changes is available only
indirectly  through changes in the time derivatives of the observed
state  variables. Even modest levels of observation noise are
sufficient  to confound this attempt. We conclude that higher
frequency  observations, and /or lower observation noise, would be
required  to learn about temporal variation in these community
properties  from observations of state variables alone. 

A twin experiment with similar forcing, sampling pattern and
observation noise  to the historical OSP observations yielded
qualitatively  equivalent results to those obtained using the real
data.  While not conclusive, this does suggest that the limited
information  about state and parameters obtained using the historical
observations may be attributed to their sparseness and high
observation  noise, rather than an inconsistency with structural model
assumptions. 

It is common practice to distinguish short-term
forecasts,  in which uncertainty is dominated by the error in
estimates  of the current system state, from longer-term forecasts or
projections,  in which uncertainty may be dominated by errors in model
structure, errors in parameter estimates, and the underlying stochastic
process error.  The methods used here allow us to move seamlessly from
short-term  to long-term forecasts. The forecast results are
encouraging (Figure \ref{fig:OSP_state_forecast}),  especially given that the inference procedure
and  observations have provided information about a small subset only
of  model parameters. This limited information, combined with prior
information  on other parameters, is sufficient to produce a long-term
forecast  that agrees both qualitatively and quantitatively with observations. 

Given the limited identifiability of both parameters and the related
stochastic biogeochemical processes, one could reasonably ask whether the
model is over-parameterised. This would be the case if we were building a
model specifically for the purpose of explanation or prediction at OSP that
ignored prior information on model structure and parameters.  However, we are
engaged in developing a generic model, based on well accepted principles and
strong prior information. The model is applied at OSP, but we envisage the
same model (or a similar model) being applied at many other locations, and in
the long run used as a basis for basin-scale or global BGC models spanning
many different environmental conditions.  Under these circumstances, it would
be inappropriate to eliminate processes from the model on the grounds that
they are not important at OSP, or that they are not identifiable from a
particular set of historical observations from OSP. We are interested rather
in the question of what such a model allows us to infer and predict about OSP
and other regions, given generic objective prior information and the limited
available observations.

Models with many, poorly identified parameters can be subject to
over-tuning and  poor predictive performance, especially if parameter
estimation  procedures are heuristic, and/or are designed to produce a
single ``optimal''  parameter set. The BHM framework and inference
procedures used  here provide protection against over-tuning. The
posterior  distribution yields samples from the full range of possible
parameters  and states, conditional on priors and observations, and
it therefore  provides a realistic picture of the effects of equifinality
\citep{VonBertalanffy1969, beven1992} on model hindcasts and
predictions. The performance of the posterior for  the long-term
forecast for OSP (Figure\ref{fig:OSP_state_forecast}) supports this conclusion.

Emerging observing systems promise much richer data sets than in the
past.  New automated in situ and remote sensors can provide data for
more  variables with much higher temporal resolution and/or spatial
coverage.  The twin experiment with daily observations presented here
provides  a hint of what we might expect from such improved observing
systems.  Data assimilating models are increasingly being used to
assess the  information value of alternative observing system designs,
as part of  so-called Observing System Simulation Experiments \citep[OSSEs;
e.g.,][]{Masutani2010}.  The twin experiments presented were
intended primarily  as a check on the consistency and performance of
the inference  methods; an OSSE would require careful attention to
observing  system elements and costs, and the use of replicate
experiments.  We anticipate using the BHM framework to build OSSEs. Oceanographic field studies often include local in
situ or  ship-board experiments that effectively measure the
instantaneous  values of community ecophysiological properties. The
model  formulation proposed here offers the opportunity to integrate
these  measurements with standard observations of state variables
(biomass)  within a consistent and rigorous inference framework. We
see this  as an interesting direction for further research using both
OSSEs  and real observations.

We recognise that, in order to fully exploit the potential for OSSEs,
and for  hindcasting and forecasting more generally, it will be
necessary to  extend our approach from the 0-D box model considered here
to  spatially resolved models, including both 1-D vertical mixing
models  \citep[e.g.][]{Mattern2010} and 3-D circulation models  
\citep[cf.][]{Gregg2008}.  The adoption of spatially resolved models would avoid
the  ambiguity about spatial scales inherent in the box model and
allow a  more rigorous treatment of spatial sampling errors. We do not
foresee major conceptual problems in extending the formulation to
spatially resolved models, but  Bayesian inference in these models
will involve  formidable computational challenges, and may require the
development  of effective approximate techniques.   

We believe that the example presented here delivers at least in part
on  the promise described by \cite{Berliner2003} and \cite{Cressie2009}
of  BHM as a self-consistent probabilistic framework that integrates
statistical  and mechanistic process models. The specific process
model  developed here shows promise as a basis for applications for
many  local and regional aquatic BGC applications. We hope
that  some of the methods developed here, including the use of
stochastic  processes for aggregate community properties, will
find  broader application.

\section*{Acknowledgements}
We are indebted to two anonymous referees for their careful and constructive
reviews of an earlier version of this paper, which led to significant
improvements both in clarity and presentation. The guidance provided by
Editor-in-Chief David Schimel is gratefully acknowledged. We are grateful for
helpful comments and constructive criticism at various stages of this research
from John Taylor, Richard Matear, Yong Song, Nugzar Margvelashvili and David
Clifford. We are also grateful for the input of many colleagues in helping us
formulate our thinking, and in particular we thank Karen Wild-Allen for her time
and patience in helping the more statistically-inclined members of the project
team understand more of BGC modelling and its applications. We have
benefited from Peter Oke's many insights from a data assimilation
perspective. We thank Arnaud Doucet for his insights on sequential Monte
Carlo, which led to a much-improved inference algorithm. Bronwyn Harch was
instrumental in starting the project leading to the work reported here, and we
thank her for her enthusiasm and support. Finally, the third author is
particularly grateful to Mavis Dias for her help, enthusiastically given, at
a number stages in the preparation of this paper.

\appendix

\section{Process Model: Supplementary Material}

\subsection{A simple adaptive model of phytoplankton growth and composition in response to light, nutrient, and temperature.}\label{app:Pg_derivation}

The phytoplankton growth model used in this paper predicts changes in
phytoplankton specific growth rate $g$ and composition
(nitrogen:carbon chlorophyll-a:carbon ratios) in response to changes
in incident irradiance $E$, temperature $T$, and dissolved inorganic
nitrogen $N$.  The formulation represents a compromise between realism
and complexity. Consistent with the BHM framework, we have sought a
formulation  that explicitly connects the key observed quantity
chlorophyll-a ($Chla$)  to the state variable of phytoplankton biomass
($P$)  using the common currency of nitrogen. By explicitly treating
changes in nitrogen:carbon ratios as well, the formulation would also
support observations of dissolved oxygen or dissolved inorganic
carbon, although we do not treat those in this paper. At the same
time, we have avoided introducing additional hidden state variables,
and we have sought to minimize the number of new parameters.

This formulation draws on the adaptive phytoplankton growth model of
\citet{Geider1998},  who specified the carbon-specific phytoplankton
growth  rate, $g_C$, and the  nitrogen-specific phytoplankton growth
rate, $g_N$, as follows:

\begin{equation}
g_C = g^m_C . (1 - \exp(-\alpha.\lambda.E / g^m_C) ),
\label{eq:gc}
\end{equation}

\begin{equation}
g^m_C = g^{max} . Tc . (\chi-\chi^{min}) / (\chi^{max} - \chi^{min} ),
\end{equation}

\begin{equation}
g_N = V^m . N / (K + N),
\end{equation}

\begin{equation}
V^m = V^{max}. Tc . (\chi^{max} - \chi) / (\chi^{max} - \chi^{min}).
\end{equation}

Here, the temperature correction factor $Tc$ is based on a ``$Q_{10}$
factor'' (see Eq. (\ref{eq:Tc})
in the main text), $\lambda$ is the chlorophyll-a:carbon ratio, $K$ is the half-saturation constant for phytoplankton growth on $N$, and $\alpha$ is the initial slope of the photosynthesis versus irradiance curve. In \citet{Geider1998}'s model, the light-saturated $C$-specific phytoplankton growth rate, $g^m_C$, and the nutrient-saturated phytoplankton nitrogen-specific growth rate, $V^m$, depend on the nitrogen:carbon ratio, $\chi$, in a way that ensures that the ratio lies between $\chi^{min}$ and $\chi^{max}$. We may think of this formulation as an extensive version of a cell quota model. 

\citet{Geider1998} introduced a third expression for the phytoplankton $Chla$-specific growth rate, $g_{Chl}$, as a function of $g_C$, $g_N$, $\lambda$, $E$, and $\chi$. Here, we have avoided introducing new dynamic state variables for phytoplankton carbon and phytoplankton chlorophyll-a. Instead, we have sought a solution for $g_N$, $\lambda$ and $\chi$, as functions of $T$, $N$ and $E$, under conditions of balanced growth when $g_C = g_N = g_{Chla}$. It is not possible to derive an explicit expression for these solutions in the original model. However, it is possible to do so with the following key simplification. 

\citet{Behrenfeld2005} argue that phytoplankton adjust their chlorophyll-a:carbon ratio, $\lambda$, so that it is proportional to $g^m_C$ under balanced growth. We assume that, at steady-state, 

\begin{equation}
\lambda = \lambda^{max}. g^m_C / g^{max},
\label{eq:lambda}
\end{equation}

and substituting (\ref{eq:lambda}) into (\ref{eq:gc}) gives

\begin{equation}
g_C = g^m_C . (1 - \exp(-\alpha . \lambda^{max}. E / g^{max}) ). 
\end{equation}

We define $g^*_E$ as the maximum $C$-specific phytoplankton growth rate for given irradiance $E$ and temperature $T$. This is achieved at $\chi = \chi^{max}$ and is given by: 

\begin{equation}
g^*_E = g^{max} . Tc. (1 - \exp(-\alpha.\lambda^{max}. E / g^{max}) ).
\label{eq:g*E}
\end{equation}

We define $g^*_N$ as the maximum $N$-specific phytoplankton growth rate for given nutrient concentration $N$ and temperature $T$. This is achieved at $\chi = \chi^{min}$ and is given by:

\begin{equation}
g^*_N = V^{max}.Tc.N/(K+N).
\label{eq:g*N}
\end{equation}

\noindent Now, $g_C = g^*_E. (\chi-\chi^{min}) / (\chi^{max} - \chi^{min})$, and $g_N = g^*_N. (\chi^{max} - \chi) / (\chi^{max} - \chi^{min} )$.

Under balanced growth, the specific growth rate $g = g_C = g_N$, and solving for $\chi$ gives:

\begin{equation}
\chi = (g^*_E.\chi^{min} + g^*_N . \chi^{max}) / (g^*_E + g^*_N).
\label{eq:chi_app}
\end{equation} 

\noindent Substituting (\ref{eq:chi_app}) into $g_C = g^*_E. (\chi-\chi^{min}) / (\chi^{max} - \chi^{min})$, and appealing to the equality $g = g_C = g_N$, yields:

\begin{equation}
g = g^*_E.g^*_N/(g^*_E + g^*_N)
\end{equation}

\noindent and, from, (\ref{eq:lambda})

\begin{equation}
\lambda = \lambda^{max}.Tc.g^*_N /( g^*_E + g^*_N).
\end{equation}

These expressions have a simple and logical interpretation. The rates $g^*_E$ and $g^*_N$ are the potential specific phytoplankton growth rates determined by light and nutrient separately, and we can think of them as measures of light and nutrient availability as perceived by the cell. The achieved growth rate $g$ is a compromise between the two. If one is much larger than the other, then the achieved growth rate is very close to the smaller rate. Note that simple non-adaptive models of phytoplankton growth often multiply a light-limitation and a nutrient-limitation term. It is well known that this underestimates growth rates. Our approach partly avoids the defect of multiplicative growth models. In (\ref{eq:chi_app}), the nitrogen:carbon ratio approaches $\chi^{min}$ when light is much more available than nutrient ($g^*_E >> g^*_N$), and it approaches $\chi^{max}$ when $g^*_N >> g^*_E$, as one might expect. The chlorophyll-a:carbon ratio, $\lambda$, approaches $\lambda^{max}$ when light is limiting and nutrient is abundant ($g^*_N >> g^*_E$), and it approaches zero when nutrient is strongly limiting and light is saturating.  

It is possible to treat $g_N$ as an $N$-specific uptake rate, and for the maximum specific uptake rate of nitrogen ($V^{max}$) to be substantially larger than the maximum $C$-specific growth rate ($g^{max}$). This would allow rapid uptake of nitrogen in a dynamic quota model. However, given that we are considering only balanced growth here, we treat $g_N$ as a specific growth rate and assume $V^{max} = g^{max}$. We must then interpret $K$ as a half-saturation constant for growth rather than uptake: 

\begin{equation}
 K = \frac{g^{max}.T_C}{a_N}.
\label{eq:K}
\end{equation}

\noindent With this simplification (substitute (\ref{eq:K}) into the $N/(K+N)$ term of (\ref{eq:g*N})), we can write:

\begin{equation}
h_N = N/((g^{max}.Tc/a_N)+N),
\end{equation}

\noindent and from (\ref{eq:g*E}) we set

\begin{equation}
h_E = 1 - \exp(-\alpha.\lambda^{max}. E / g^{max}).
\end{equation}

\noindent As before, under the assumption of balanced growth ($g=g_C=g_N$), we obtain

\begin{equation}
g = Tc.g^{max}.h_E.h_N/(h_E+h_N),
\end{equation}

\noindent and

\begin{equation}
\chi = (h_E.\chi^{min} + h_N . \chi^{max}) / (h_E + h_N).
\end{equation}

\noindent If we set $R_N$ = $\chi^{min}/\chi^{max}$, then

\begin{equation}
\chi = \chi^{max} . (h_E.R_N + h_N) / (h_E + h_N),
\end{equation}

\noindent and

\begin{equation}
\lambda = \lambda^{max}.Tc.h_N /(h_E + h_N).
\end{equation}

\noindent The phytoplankton chlorophyll-a:nitrogen ratio equals $\lambda/\chi$, and it is given by:

\begin{equation}
Chla:N = (\lambda^{max}.Tc/\chi^{max}). h_N/(h_E.R_N + h_N). 
\end{equation}

\subsection{The transport operator}\label{app:sourcesink}
In this section we describe the transport operator, ${\bf T}({\bf
  c},{\bf x},t)$, used in equation (\ref{eq:partialc})
in the main text. The specific form applied to each of the state variables differs based on the characteristics of the specific state variable being operated on.

The change in mixed layer depth ($MLD$) is given by:
\begin{equation}
 \psi(t)=\frac{d(MLD)}{dt} \label{eq:psi},
\end{equation}

\noindent and we define

\begin{equation}
\psi^+(t) \equiv max\left\{\psi(t),0\right\}.
\end{equation}

\noindent This form of exchange across the mixed-layer has been adopted from \citet{Evans1985}. Therefore the state equations, including the effects of changes in the mixed layer, can be written as: 

\begin{equation}
	\frac{dP}{dt}=g\cdot P-gr\cdot Z + \frac{\kappa + \psi^+(t)}{MLD} \cdot (BCP-P),
\end{equation}

\begin{equation}
	\frac{dZ}{dt}=E_Z\cdot gr\cdot Z - m\cdot Z + \frac{\psi(t)}{MLD} \cdot (BCZ-Z),
\end{equation}

\begin{equation}
	\frac{dD}{dt}=(1 - E_Z)\cdot f_D\cdot gr\cdot Z + m\cdot Z - r\cdot D - S_D\cdot D / MLD + \frac{\kappa + \psi^+(t)}{MLD} \cdot (BCD-D),
\end{equation}

\begin{equation}
	\frac{dN}{dt}=-g\cdot P + (1 - E_Z)\cdot (1- f_D)\cdot gr\cdot Z + r\cdot D + \frac{\kappa + \psi^+(t)}{MLD} \cdot (BCN - N),
\end{equation}

\noindent where $\kappa$ is the background mixing, $S_D$ is the detrital sinking rate (a parameter subject to inference), and $BCP$, $BCZ$, $BCD$ and $BCN$ are the boundary conditions for $P,Z,D$ and $N$, respectively. With the exception of $BCN$ all other state variable boundary conditions are set to 0. It is assumed that as the mixed layer shoals, $P,D$ and $N$ are lost from the mixed layer, whereas $Z$ is assumed to be retained in the mixed layer. As the mixed layer deepens, $P,Z$ and $D$ concentrations will be diluted, whereas $N$ will be added in an amount proportional to the prescribed boundary condition $BCN$. The variables $\kappa,\psi,MLD$ and $BCN$ are considered to be exogenous forcing and are prescribed.

\subsection{The light model}\label{app:lightmodel}

\begin{equation}\label{eq:E}
E = E_0 . (1 - \exp(-Kz))/Kz,
\end{equation}

\noindent where $E_0$ is the mean daily photosynthetically available radiation (PAR) just below the air-sea interface, and $Kz$ is given by:

\begin{equation}
Kz = (K_W + a_{Ch} \cdot Chla) \cdot MLD.  \label{eq:az}
\end{equation}

\noindent In (\ref{eq:az}), $K_W$ is attenuation due to the seawater, $a_{Ch}$ is attenuation due to $Chla$ and $MLD$ is the mixed layer depth.

\subsection{The parameter (prior) model}\label{app:prior_model}

Where possible, the priors for the process-model parameters have been based on information in the literature. The parameters can be divided into three classes. 

For some physiological parameters, there are existing meta-analyses in
the literature that provide estimates of parameter means and
variances. Previous studies of phytoplankton by \citet{Tang1995} and
\citet{Montagnes1994}, and of zooplankton by  \citet{Hansen1997}, have
derived allometric relationships (log-log regressions) for
phytoplankton and zooplankton parameters $g^{max}$, $\lambda^{max}$,
$E_Z$, $Cl_Z$ and $I_Z$ as a function of individual size. From these
data sets, we have derived prior means and coefficients of variation
(Table \ref{tab:prior}), assuming that the phytoplankton community at
OSP is dominated by small cells (mean cell volume 100 $\mu$m$^3$), and
grazing is dominated by microzooplankton (mean individual volume $10^5
\mu$m$^3$). We have used normal prior distributions for those parameters having small
coefficient of variation, and log-normal prior distributions for the rest.

For some parameters, we can draw on a range of quoted values in the literature that are sufficient to provide crude estimates of prior mean and variance. Historical observations of the light attenuation due to water, $K_W$, and the specific absorption coefficient for chlorophyll-a, $a_{Ch}$, were taken from \citet{Kirk1994}. The maximum quantum yield is assumed constant at 0.1 mol $C$ mol photons$^{-1}$, or 1200 mg $C$ mol photons$^{-1}$. Estimates of the affinity of phytoplankton for dissolved inorganic nitrogen, $a_N$, are based on data in \citet{Hein1995} and on theoretical calculations of the diffusion limit to uptake. 

Other parameters can be regarded as semi-empirical, representing ecosystem properties and processes that are only crudely represented in the model. We do not model zooplankton respiration explicitly, and we assume that approximately half of the ingested nitrogen that does not appear as an increase in biomass is released as unassimilated fecal pellets, and half is lost through respiration and excretion of dissolved inorganic nitrogen \citep[cf.][]{Parsons1973}, so $f_D$ is given a prior mean of 0.5 and a small CV of 0.1 (Table \ref{tab:prior}). Because zooplankton grazing is assumed to be dominated by micro-zooplankton, we have assigned the detrital sinking rate $S_D$ a relatively small prior mean of 5 m d$^{-1}$, with a large CV of 1.0. 

Detrital organic matter comprises diverse organic compounds that vary widely in their susceptibility to bacterial attack and remineralization. In the model, very labile organic nitrogen compounds such as amino acids, which may be utilized and remineralized on time scales of hours, are treated implicitly as part of the fraction $(1-E_Z).(1-f_D)$  of ingestion that is released directly as dissolved inorganic nitrogen. A mean remineralization rate of 0.1 d$^{-1}$ is applied to the remaining detritus, with a relatively large CV of 0.5. The quadratic mortality rate for zooplankton, $m_Q$, is an empirical ecosystem parameter, representing the density-dependent predation on micro-zooplankton. It has been assigned a mean value of 0.01 d$^{-1}$ (mg $N$ m$^{-3}$)$^{-1}$, corresponding to a mortality rate of 0.1 d$^{-1}$ at typical micro-zooplankton biomass levels of 10 mg $N$ m$^{-3}$. Because we have little prior information to constrain $m_Q$, we have assigned it a large CV of 1.0.

The scale factors $PDF$ and $ZDF$ have been assigned a prior mean of 0.2. This is a relatively low value, and it corresponds to a diverse community in which the community mean properties show substantially less variation than those of individual species (see Appendix \ref{app:ATLG}). The prior variance is set to 0.4, so the prior distribution allows higher values of $PDF$ and $ZDF$, and also less diverse communities.

\begin{table}[tp]
\centering
\caption{The priors on parameters used in the stochastic NPZD model
  are all log-normal, with mean ($\mu$) and standard deviation
  ($\sigma$) on the log-scale, except for $S_D$ which is Gaussian.}
\begin{tabular}{llll}\hline
Parameter&Description&Mean($\mu$)&SD ($\sigma$)\\
\hline
$K_W$&Light Attenuation: Water&0.03 m$^{-1}$&0.2\\
$a_{Ch}$&Light Attenuation: Chla&0.04 m$^2$ mg Chla$^{-1}$&0.3\\
$S_D$&Detrital Sinking Rate&5 m d$^{-1}$&1.0\\
$f_D$&Fraction of grazing to detritus&0.5&0.1\\
\hline
$PDF$&Phytoplankton Diversity Factor&0.15&0.4\\
$ZDF$&Zooplankton Diversity Factor&0.15&0.4\\
$\mu_{g^{max}}$&Maximum Carbon Specific Growth Rate&1.2 d$^{-1}$&0.63\\
$\mu_{R_N}$&Ratio between $\chi^{min}$ and $\chi^{max}$&0.25&0.3\\
$\mu_{\lambda^{max}}$&Maximum $Chla$ to $C$ratio&0.03&0.37\\
$\mu_{a_N}$&Phytoplankton affinity for $N$&0.3 d$^{-1}$ mg $N^{-1}$ m$^3$&1\\
$\mu_{I_Z}$&Maximum Zooplankton ingestion rate&4.7 d$^{-1}$&0.7\\
$\mu_{Cl_Z}$&Maximum Zooplankton clearance rate&0.2 m$^3$ mg $N^{-1}$ d$^{-1}$&1.3\\
$\mu_{E_Z}$&Zooplankton growth efficiency&0.32&0.25\\
$\mu_{m_Q}$&Zooplankton quadratic mortality rate&0.01 d$^{-1}$ mg $N^{-1}$ m$^3$&1\\
$\mu_{r_D}$&Detrital remineralization rate&0.1 d$^{-1}$&0.5\\
\hline
\end{tabular}
\label{tab:prior}
\end{table}

\section{Autoregressive Process with Log-Normal Innovations}\label{app:ATLG}

Suppose that a time series $\{B(t)\}$ evolves according to an AR(1)
process with an intrinsic time-scale of $\tau$. Then, from equation
(20) 
 in the main text,

\begin{equation}\label{eqn:AR}
B(t+1)=B(t)(1-1/\tau)+\zeta_B(t)/\tau.
\end{equation}

\noindent The innovation sequence $\{\zeta_B(t)\}$ is independent of
$\{B(t)\}$  and  is assumed to follow a log-normal distribution, so that
$Z=\log(\zeta_B(t)) \sim N(\mu_z,\sigma_z^2)$.  The moments  of the
log-normal distribution are  immediately  available from those of the
normal  distribution since  $ E[\zeta_B(t)^s] = E[ \exp(sZ)]$, the
moment generating function of the underlying normal distribution. We find then that
 
 \begin{equation}
 E(\zeta_B(t)^s) = \exp( s\mu_z + s^2\sigma_z^2/2), \quad s \ge 0. 
 \label{eqn:moments}
 \end{equation}

\noindent Hence the mean and variance of $\{\zeta_B(t)\}$ follow directly as

\begin{eqnarray}
\mu_\zeta & = & \exp(\mu_z+ \sigma_z^2/2) \nonumber , \\
\sigma^2_\zeta & = & \exp(2\mu_z)(\exp(2\sigma_z^2)-\exp(\sigma_z^2)).
\end{eqnarray}

\noindent The coefficient of variation, $CV(\zeta_B) = \sigma_\zeta/\mu_\zeta $, is given by

\begin{equation}
CV(\zeta_B)=(\exp(\sigma_z^2)-1)^{1/2},
\end{equation}

\noindent and so it does not depend on the mean.

Assuming second-order stationarity of $\{B(t)\}$ requires that $\vert
1- 1/\tau \vert < 1$, so $\tau>1/2$. Taking expectations on
both sides of  equation (\ref{eqn:AR}), we find (using an obvious notation):

\begin{equation}
\mu_B=\mu_B(1-1/\tau)+\mu_\zeta/\tau,
\end{equation} 

\noindent so

\begin{equation}
\mu_B = \mu_\zeta = \exp \left( \mu_z+\sigma_z^2/2 \right).
\label{eqn:MUB}
\end{equation}

\noindent Next, calculating the variance of both sides using
the independence assumption above, we obtain again using an obvious notation):

\begin{eqnarray}
  \sigma^2_B &=& (1-1/\tau)^2\sigma^2_B + \sigma^2_\zeta/\tau^2  \nonumber \\
            &=& (2\tau-1)^{-1/2}\sigma^2_\zeta.
                       \end{eqnarray}

\noindent Since $\mu_B = \mu_\zeta$,

\begin{eqnarray}
CV(B)^2 &=& CV(\zeta_B)^2/(2\tau-1) \nonumber \\
       &=& \left( \exp(\sigma_z^2) - 1 \right) / (2\tau-1).
\label{eqn:CVB}
\end{eqnarray}

We may apply this result as follows. Suppose  a community is made up
of $n$ species,  each  with biomass fractions $\{p_i: i=1, \dots, n;
\sum p_i = 1 \}$,  and suppose further that the species-specific growth  rate parameters
$\boldsymbol{\phi}_i$ are log-normally distributed, where
$log(\boldsymbol{\phi}_i)$ has mean $\mu_{l \phi}$ and variance
$\sigma_{l \phi}^2$.  We are interested in the biomass-weighted
community  value, $\Phi \equiv \sum{p_i}{\boldsymbol{\phi}_i}$. Hence,

\begin{eqnarray}
E(\Phi) & = & \sum p_i E(\boldsymbol{\phi}_i) \nonumber \\
                  & = & \exp(\mu_{l\phi}+\sigma_{l\phi}^2/2).
\label{eqn:Ephi}
\end{eqnarray}

\noindent Further,

\begin{equation}
var(\Phi) =\sum{p_i}^2var(\boldsymbol{\phi}_i) 
= \sum{p_i}^2 \exp(2\mu_{l \phi} ) (\exp(2\sigma_{l \phi}^2) - \exp(\sigma_{l \phi}^2)).
\end{equation} 

\noindent Thus,

\begin{equation}\label{eqn:CVPHI}
CV(\Phi)^2  =  \sum{p_i}^2 (\exp(\sigma_{l \phi}^2) - 1).
\end{equation}

In our case, we want to choose  $\mu_\zeta$ and 
$\sigma_\zeta$ so that $E(B) = E(\Phi)$ and 
$ CV(B)^2 = CV(\Phi)^2$. Using (\ref{eqn:MUB}), (\ref{eqn:CVB}),
(\ref{eqn:Ephi}) and (\ref{eqn:CVPHI}),
we have

\begin{eqnarray}
exp(\mu_\zeta + \sigma_{\zeta}^2/2) & = & \exp(\mu_{l\phi} +
\sigma^{2}_{l\phi}/2),\\
(\exp(\sigma_\zeta^2) - 1)/(2\tau - 1) & = & \sum{p_i}^2. (\exp(\sigma_{l\phi}^2) - 1).
\end{eqnarray}

Finally,
\begin{eqnarray}
\mu_z & = & \mu_{l \phi}+\sigma_{l \phi}^2/2-\sigma_{z}^2/2, \\
\sigma_z^2 & = & \log \left( 1 + (2\tau - 1) 
\sum{p_i}^2(\exp(\sigma_{l \phi}^2) - 1) \right).
\end{eqnarray}

For those eco-physiological parameters modelled here as a stochastic
autoregressive  process $\{B(t)\}$, we derive prior estimates of
$\mu_{l\phi}$ and $\sigma_{l\phi}$  from the literature. We treat the
quantities $\sum p_i^2$ as constants. For the phytoplankton  community,
$\sum p_i^2 \equiv PDF^2$, and for the zooplankton community
$\sum p_i^2 \equiv ZDF^2$. The diversity factors $PDF$ and $ZDF$
(recalling that these scale the prescribed $\sigma_{l\phi}$ values)
and $\mu_{l\phi}$ are parameters subject to inference, and they are given prior
distributions (see Appendix \ref{app:prior_model}).

\section{Particle Markov Chain Monte Carlo Sampling}\label{app:PMMH}

For brevity, we use the notation ${\bf W}_{i:j}$ (where $i \le j$) to denote
the sequence $({\bf W}_i,\ldots,{\bf W}_j)$, and likewise for ${\bf
  Y}_{i:j}$. We may write the posterior distribution as
\begin{eqnarray}
[{\bf W}_{1:T},\boldsymbol{\theta}_{\bf W},
\boldsymbol{\theta}_{\bf{Y}}\vert {\bf Y}_{1:T}] & = & 
[\mathbf{W}_{1:T}|\boldsymbol{\theta}_{\bf W},
\boldsymbol{\theta}_{\bf Y},\mathbf{Y}_{1:T}] 
[\boldsymbol{\theta}_{\bf W},
\boldsymbol{\theta}_{\bf Y}|\mathbf{Y}_{1:T}].\label{eqn:pmmh-factor}
\end{eqnarray}
The first factor of Equation (\ref{eqn:pmmh-factor})  may be computed using a particle  filter~\citep{Gordon1993,Doucet2001}, with  parameters fixed according to a draw from the  second factor, considered below. At time $t$,  let $[\mathbf{W}_{1:t} | \boldsymbol{\theta}_{\bf W}, \boldsymbol{\theta}_{\bf Y},\mathbf{Y}_{1:t}]$ be  represented by a set of $N$ samples (particles)  $\mathbf{w}^i_{1:t}$ with associated weights  $\alpha^i_t$, for $i = 1,\ldots,N$. Given  $\{(\mathbf{w}^i_{1:t-1},\alpha_{t-1}^i)\}$,  we may recursively obtain $\{(\mathbf{w}^i_{1:t}, \alpha_t^i)\}$ by setting, for $i = 1,\ldots,N$,

\begin{eqnarray}
\mathbf{w}^i_t &\sim& [ \mathbf{W}_t | 
\mathbf{W}_{t-1} = \mathbf{w}^i_{t-1}, 
\boldsymbol{\theta}_{\bf W}] \\
\alpha^i_t &=& \alpha^i_{t-1} \cdot [\mathbf{Y}_t|\mathbf{W}_t 
= \mathbf{w}^i_t, \boldsymbol{\theta}_{\bf Y}].
\label{eq:walpha}
\end{eqnarray}
The procedure is initialised with $\mathbf{w}_0^i \sim [\mathbf{W}_0]$ and
$\alpha_0^i = 1$, for $i = 1,\ldots,N$. After several iterations, the sample tends to degenerate to a situation where only a single particle has
significant weight. For this reason, resampling is typically performed after
each step, redrawing particles with probability proportional to weights, and
resetting all weights to be identically one. This has the effect of
eliminating particles of low weight and replicating particles of high
weight~\citep{Gordon1993,Kitagawa1996}.

Note that, by time $T$, the procedure has produced  weighted samples of the target factor  $[\mathbf{W}_{1:T}|\boldsymbol{\theta}_{\bf W}, \boldsymbol{\theta}_{\bf Y},\mathbf{Y}_{1:T}]$  of Equation (\ref{eqn:pmmh-factor}).  Sampling-importance-resampling (SIR) may be  used to obtain a single sample from among these,  making a single multinomial draw from the set  $\{\mathbf{w}^i_{1:T}\}$ with assigned probabilities  $\{\alpha_T^i/\sum_{j=1}^N \alpha_T^j\}$.

Turning to the second factor of Equation (\ref{eqn:pmmh-factor}), note that
\begin{eqnarray}
[\boldsymbol{\theta}_{\bf W},\boldsymbol{\theta}_{\bf Y}
|\mathbf{Y}_{1:T}] &\propto& [\mathbf{Y}_{1:T}|
\boldsymbol{\theta}_{\bf W},\boldsymbol{\theta}_{\bf Y}]
[\boldsymbol{\theta}_{\bf W},\boldsymbol{\theta}_{\bf Y}],
\label{eq:thetaW}
\end{eqnarray}
and that a particle filter may again be used to compute  the likelihood term:
\begin{eqnarray}
[\mathbf{Y}_{1:T}|\boldsymbol{\theta}_{\bf W},
\boldsymbol{\theta}_{\bf Y}] &=& [\mathbf{Y}_1|
\boldsymbol{\theta}_{\bf W},\boldsymbol{\theta}_{\bf Y}] 
\prod_{t=2}^T [\mathbf{Y}_t|\mathbf{Y}_{1:t-1},
\boldsymbol{\theta}_{\bf W},\boldsymbol{\theta}_{\bf Y}] \\
&=& \int_{-\infty}^{+\infty} [\mathbf{Y}_1|
\mathbf{W}_1,\boldsymbol{\theta}_{\bf Y}]
[\mathbf{W}_1]d\mathbf{W}_1 \times \notag\\
&& \prod_{t=2}^T \int_{-\infty}^{+\infty} 
[\mathbf{Y}_t|\mathbf{W}_t,\boldsymbol{\theta}_{\bf Y}]
[\mathbf{W}_t|\mathbf{Y}_{1:t-1},
\boldsymbol{\theta}_{\bf W},\boldsymbol{\theta}_{\bf Y}]
d\mathbf{W}_t \\
&\approx& \prod_{t=1}^T \frac{1}{N} \sum_{i=1}^N 
\alpha_t^i,\label{eqn:pmmh-likelihood}
\end{eqnarray}
where we have assumed in arriving at the last line  that resampling is performed at each recursion of  the particle filter. See \citet{Andrieu2010} for details.

The computation of this likelihood facilitates a Metropolis-Hastings
\citep{Metropolis1953, Hastings1970} sampling of $[\boldsymbol{\theta}_{\bf W}, \boldsymbol{\theta}_{\bf Y}|\mathbf{Y}_{1:T}]$. For each parameter sample accepted, a state sample is drawn from the same particle filter used to obtain a likelihood estimate, resulting in a full sample of the posterior distribution given by Equation (\ref{eqn:pmmh-factor}).

Finally, we require a suitable proposal distribution  for the Metropolis-Hastings sampler. In this work,  we use the approximate (log-)normal posterior  furnished by a joint unscented Kalman  filter~\citep{Julier1997,Wan2000} performing online process and parameter inference. After some number  of steps, this is discarded in favour of a  (log-)normal proposal adapted from the history  of posterior samples~\citep{Haario2001}.

\section{Case Study: Supplementary Material}\label{app:CS_SUP}

\subsection{Forcing Data}
\begin{figure}[tp]
	\centering
		\includegraphics[width=1.0\textwidth]{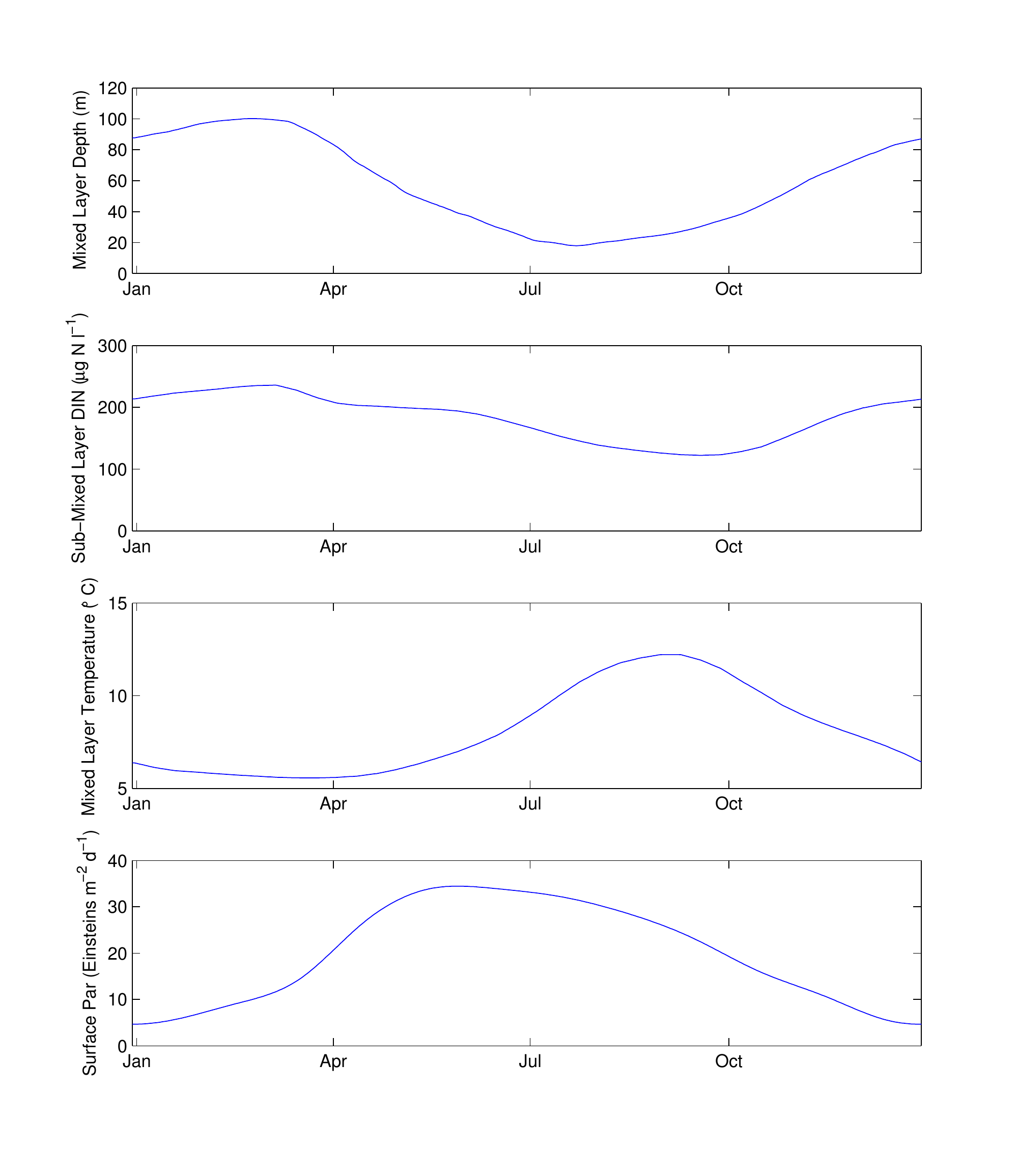}
	\caption{The climatological forcing obtained from \citet{Matear1995} used in the twin experiment.}
	\label{fig:TEforcing}
\end{figure}

\begin{figure}[tp]
	\centering
		\includegraphics[width=1.0\textwidth]{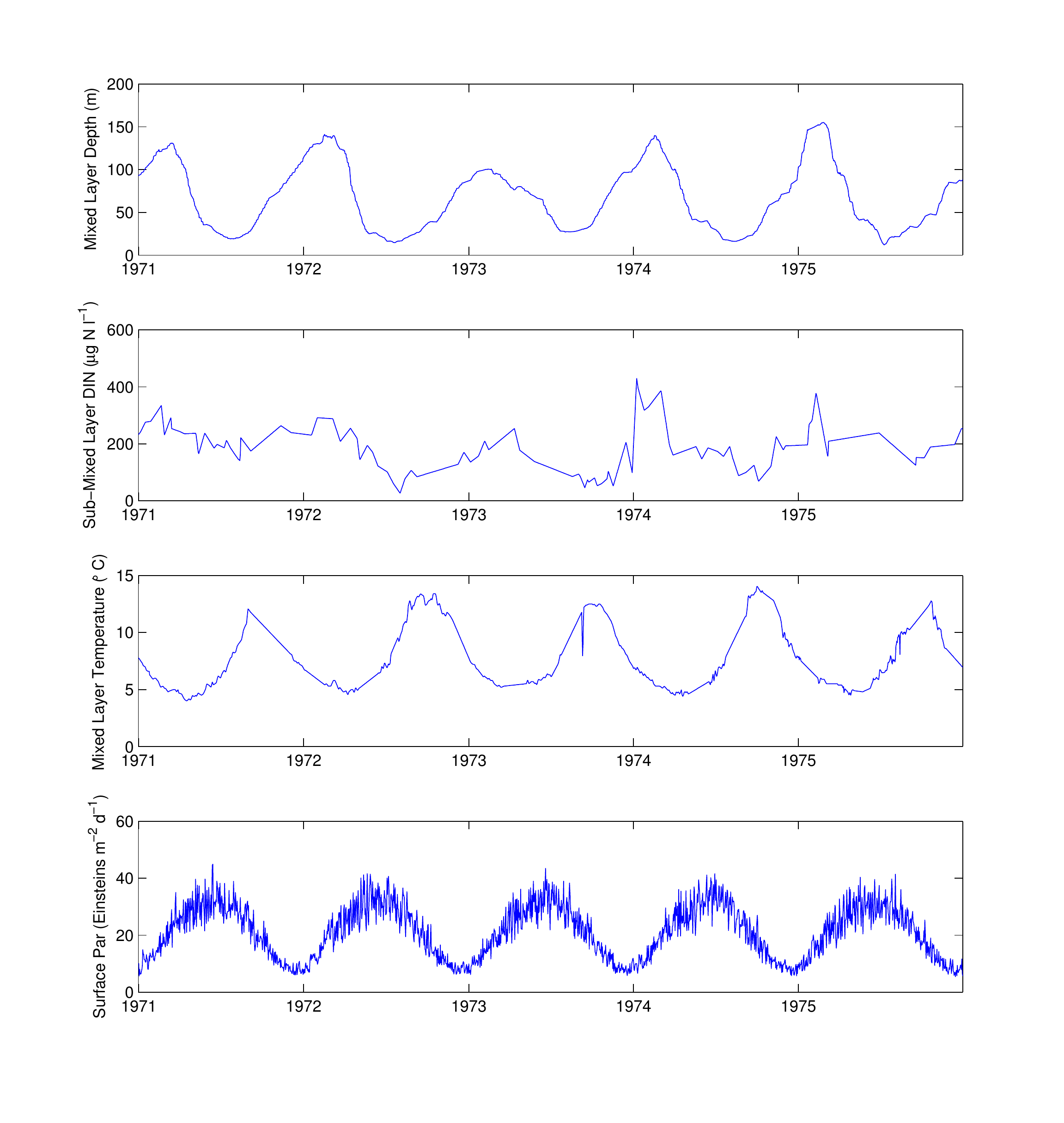}
	\caption{The forcing used in the Ocean Station Papa case study.}
	\label{fig:OSPforcing}
\end{figure}

The forcing data used for the climatological twin-experiment and
case-study was derived from that used in the study outlined in
\citet{Matear1995}. The exogenous forcing that was applied is given in
Figure \ref{fig:TEforcing}. The forcing used for the 1971-1975 OSP
experiment using the historical observations was again derived from
the dataset used by \citet{Matear1995} and is contained in Figure
\ref{fig:OSPforcing}. The time series of Mixed Layer Depth (MLD) is
used in equation (\ref{eq:partialc}) 
in the main text to form the transport operator and is
discussed in Appendix \ref{app:sourcesink}. Furthermore, the daily
change in MLD (\ref{eq:psi}) is used to parameterise the exchange
across the base of the mixed layer. The sub-Mixed Layer DIN is the
boundary condition (BCN) referred to in Appendix
\ref{app:sourcesink}. The Mixed Layer Temperature is used in the
temperature correction term (equation (\ref{eq:Tc}) in the main
text)
for many of the rate
processes given in Section \ref{sec:procmodel} 
in the main text and Appendix
\ref{app:Pg_derivation}, while the photosynthetic available radiation
(PAR) immediately below the surface is used to drive the
photosynthesis in the phytoplankton growth model (Appendix
\ref{app:Pg_derivation}).

\clearpage



\end{document}